\documentclass{aastex}
\usepackage{emulateapj5}

\submitted{Submitted 24 September 2001; Accepted 4 December 2001 -- To appear in the March, 2001 AJ}
\newcommand{\etal}{{\it et\thinspace al.}\ }
\newcommand{\kms}{km\thinspace s$^{-1}$}
\newcommand{\simlt}{\ {\raise-.5ex\hbox{$\buildrel<\over\sim$}}\ }

\lefthead{Salzer \etal}
\righthead{KPNO International Spectroscopic Survey}

\begin{document}

\title{The KPNO International Spectroscopic Survey.  \\III. [\ion{O}{3}]-selected Survey List.}

\author{John J. Salzer\altaffilmark{1}, Caryl Gronwall\altaffilmark{1,2}, and Vicki L. Sarajedini\altaffilmark{3}}
\affil{Astronomy Department, Wesleyan University, Middletown, CT 06459; 
slaz@astro.wesleyan.edu}

\author{Valentin A. Lipovetsky\altaffilmark{1,4} and Alexei Kniazev\altaffilmark{1,5}}
\affil{Special Astrophysical Observatory, Russian Academy of Sciences, Nizhny Arkhyz, Karachai-Circessia 357147, Russia}

\author{J. Ward Moody}
\affil{Department of Physics \& Astronomy, Brigham Young University, Provo, UT 84602; jmoody@astro.byu.edu}

\author{Todd A. Boroson}
\affil{National Optical Astronomy Obs., P.O. Box 26732, Tucson, AZ 85726; tyb@noao.edu}

\author{Trinh X. Thuan}
\affil{Astronomy Department, University of Virginia, Charlottesville, VA 22903; txt@starburst.astro.virginia.edu}

\author{Yuri I. Izotov}
\affil{Main Astronomical Observatory, National Academy of Sciences of Ukraine, Goloseevo, Kiev 03680, Ukraine; izotov@mao.kiev.ua}

\author{Jos\'e L. Herrero}
\affil{BBN Technologies, Cambridge, MA 02140; jose@world.std.com}

\author{Lisa M. Frattare\altaffilmark{1}}
\affil{Space Telescope Science Institute, Baltimore, MD 21218; frattare@stsci.edu}

\altaffiltext{1}{Visiting Astronomer, Kitt Peak National Observatory. 
KPNO is operated by AURA, Inc.\ under contract to the National Science
Foundation.} 
\altaffiltext{2}{present address: Department of Physics \& Astronomy, Johns Hopkins University,
Baltimore, MD 21218; caryl@adcam.pha.jhu.edu.}
\altaffiltext{3}{present address: Department of Astronomy, University of Florida,
Gainesville, FL 32611; vicki@astro.ufl.edu.}
\altaffiltext{4}{Deceased 22 September 1996.} 
\altaffiltext{5}{present address: Max Planck Institut f\"ur Astronomie, K\"onigstuhl 17, D-69117,
Heidelberg, Germany; kniazev@mpia.de.}

%\clearpage

\begin{abstract}
The KPNO International Spectroscopic Survey (KISS) is an objective-prism
survey for extragalactic emission-line objects.  It combines many of the 
features of previous slitless spectroscopic surveys with the advantages of 
modern CCD detectors, and is the first purely digital objective-prism survey
for emission-line galaxies.  Here we present the first list of emission-line 
galaxy candidates selected from our blue spectral data, which cover the wavelength
range 4800 -- 5500 \AA.  In most cases, the detected emission line is 
[\ion{O}{3}]$\lambda$5007.  The current survey list covers a one-degree-wide strip 
located at $\delta$ = 29$\arcdeg$~30$\arcmin$ (B1950.0) and spanning the right
ascension range 8$^h$~30$^m$ to 17$^h$~0$^m$.  An area of 116.6 deg$^2$ is covered.  
A total of 223 candidate emission-line objects have been selected for inclusion in the 
survey list (1.91 deg$^{-2}$).  We tabulate accurate coordinates and photometry 
for each source, as well as estimates of the redshift, emission-line flux, and 
equivalent width based on measurements of the digital objective-prism spectra.  
The median apparent magnitude of the sample is B = 18.2, and galaxies with 
redshifts approaching z = 0.1 are detected.  The properties of the KISS 
emission-line galaxies are examined using the available observational data, and 
compared to previous surveys carried out with photographic plates as well as with 
the H$\alpha$-selected portion of KISS.
\end{abstract}

% The different journals have different requirements for keywords.  The
% keywords.apj file, found on aas.org in the pubs/aastex-misc directory, 
% contains a list of keywords used with the ApJ and Letters.  These are 
% usually assigned by the editor, but authors may include them in their 
% manuscripts if they wish. 


%************************************************************************

\section{Introduction}

The KPNO International Spectroscopic Survey (KISS) is an ongoing objective-prism
survey which targets the detection of large numbers of extragalactic emission-line
sources.  KISS attempts to build upon previous, extremely fruitful surveys for
these types of objects, which have been responsible for the cataloging of a large
proportion of the known starburst galaxies, Seyfert galaxies, and QSOs.  Our
survey method is similar to many of these previous surveys, which have been 
carried out with Schmidt telescopes and photographic plates (e.g., Markarian 1967, 
Smith \etal 1976, MacAlpine \etal 1977, Pesch \& Sanduleak 1983, Wasilewski 1983, 
Markarian \etal 1983, Zamorano \etal 1994, Popescu \etal 1996, Surace \& Comte 
1998, Lipovetsky \etal 1998).  The fundamental difference between KISS and these 
previous surveys is that we utilize a CCD as our detector.  There are several 
obvious advantages of CCDs over photographic plates for this type of survey, including
much higher quantum efficiency, lower noise, good spectral response over the
entire optical portion of the spectrum, and large dynamic range.  In addition,
CCDs enable us to use automated selection methods to detect 
emission-line galaxies (ELGs), and allow us to quantify the selection function 
and completeness limit directly from the survey data.   With the advent of large
format CCDs in the past decade, the large areal coverage provided by the wide-field
imaging capability of Schmidt telescopes makes digital surveys like KISS possible. 
The combination of increased depth and large areal coverage leads to substantial 
improvements compared to the previous photographic surveys listed above.

The primary goal of KISS is to produce a high-quality survey whose selection
function and completeness limits can be accurately quantified so that the 
resulting catalog of ELGs will be useful for a broad range of studies
requiring statistically complete galaxy samples.  We also want to reach
substantially deeper than previous objective-prism surveys.

A complete description of the survey method employed for KISS is given
in the first paper in this series (Salzer \etal 2000, hereafter Paper I).
KISS is a {\it line-selected} survey, meaning that the objective-prism
spectra are searched for the presence of an emission feature.  As described 
in Paper I, the first survey strip was observed in two distinct spectral 
regions.  One covered the blue portion of the optical spectrum (4800 -- 5500 \AA), 
while the second covered the wavelength range 6400 -- 7200 \AA\ in the red 
part of the spectrum.  The first red survey list is given in Salzer \etal
(2001, hereafter KR1).  The current paper presents the initial KISS list
of [\ion{O}{3}]-selected ELG candidates.  The format of the current paper
follows closely that of KR1.  In addition to listing the ELGs, we 
provide substantial observational data for each object.  This includes 
accurate photometry and astrometry for each source, as well as estimates of 
each galaxy's redshift, line flux, and equivalent width.  These data are 
used to examine the properties of the KISS ELGs in Section 4.

%************************************************************************

\section{Observations}

All survey data were acquired using the 0.61-meter Burrell Schmidt 
telescope\footnote{Observations made with the Burrell Schmidt of the
Warner and Swasey Observatory, Case Western Reserve University.
During the period of time covered by the observations described 
here, the Burrell Schmidt was operated jointly by CWRU and KPNO.}. 
The detector used for all data reported here was a 2048 $\times$ 
2048 pixel STIS CCD (S2KA).  This CCD has 21-$\micron$ pixels, which 
yields an image scale of 2.03 arcsec/pixel.  The overall field-of-view 
is 69 $\times$ 69 arcmin, and each image covers 1.32 square degrees. 
Due to the coarse pixel scale, seeing variations do not adversely affect 
the survey data.  Nearly all of the direct and spectral images used to
construct the survey are undersampled (i.e., have point-spread-function
widths of less than two pixels).
The blue survey spectral data were obtained with a 2$\arcdeg$ prism,
which provided a reciprocal dispersion of 19 \AA/pixel at 5000 \AA.
The spectral data were obtained through a special filter designed for
the survey, which covered the spectral range 4800 -- 5500 \AA\ (see
Figure 1 of Paper I for the filter transmission curve).

The survey data consist of the spectral images, obtained with the 
objective prism on the telescope, plus direct images taken without
the prism through standard B and V filters.  Full details are presented
in Paper I.  The primary emission line detected in the blue spectral 
data is [\ion{O}{3}]$\lambda$5007.  It is possible that the H$\beta$
line could be stronger than [\ion{O}{3}]$\lambda$5007 and hence be the
feature seen in the survey spectra.  However, ELGs with H$\beta$
stronger than [\ion{O}{3}] tend to be fairly luminous starburst
galaxies with low equivalent width lines.  Such objects tend to be
missed in line-selected objective-prism surveys like KISS, unless the
survey is carried out in the red and selects via H$\alpha$.  Hence, we 
expect that the large majority of ELGs found in the current survey are
detected via their [\ion{O}{3}] emission.  Follow-up spectra for 123 blue-selected
KISS ELGs confirm this (see section 4).  The redshift range over which
we are able to detect ELGs, which is limited by the spectral filter
described above, is z = 0 to 0.09 for the [\ion{O}{3}] line, and
z = 0 to 0.13 for H$\beta$.  Finally, we mention the possibility
that higher redshift galaxies (z = 0.29 to 0.47) might be detected
via [\ion{O}{2}]$\lambda$3727 emission.  These objects would need to
be quite luminous and possess fairly strong emission lines in order to
be detectable.  To date no such high redshift objects have been
found in our limited follow-up spectra (although some have been
found in the red survey -- see KR1).

This blue survey consists of a contiguous strip of fields observed at a
constant declination ($\delta$ = 29$\arcdeg$~30$\arcmin$ (B1950.0)).  
The right ascension range covered is 8$^h$~30$^m$ to 17$^h$~0$^m$ (B1950.0).  
This area was chosen to overlap completely the Century Redshift Survey (Geller 
\etal 1997; Wegner \etal 2001).  Table 1 lists information about the observing 
runs during which the survey observations were obtained.  The first column 
gives the UT dates of the run,
while the second column indicates the number of nights on which observations
were obtained.  At least some data were obtained on 40 of 53 scheduled nights
(75\%).  The last two columns indicate the number of direct and spectral
images, respectively, obtained during each run.  It was common practice to 
obtain both direct and objective-prism images during each run, with the prism 
being on the telescope for about half of each block of time.   Note that the
values listed in columns 3 and 4 represent the number of images actually used
for the survey.  Due to imperfect observing conditions, many images taken
during the earlier runs were repeated later in order to replace lower quality 
images with better ones.  The most common problems plaguing the early data
were poor telescope focus and wind shaking of the telescope which resulted
in extended images.  The bulk of the data used to construct the survey were
obtained in 1996 and 1997.

%\placetable{table:tab1}

A goal of the survey was to obtain data that were of uniform depth and
image quality for all survey fields.  To that end, we employed a fixed
exposure time for all survey observations.  Direct images had exposure
times of 300 s in V and 600 s in B (although observations taken during
the first few runs were slightly longer).  
All spectral data consist of four 720 s exposures of each field, for
a total exposure time of 48 minutes per field.  The telescope was dithered 
by $\sim$10 arcsec between exposures in order to move sources off of
bad columns on the CCD.  Data processing procedures are detailed
in Paper I.  The analysis of the survey data was carried out using an
IRAF\footnote{IRAF is distributed by the National Optical Astronomy 
Observatories, which are operated by AURA, Inc.\ under cooperative 
agreement with the National Science Foundation.}-based software package 
written by members of the KISS team.  This package is described in 
Herrero \etal (2002).

%************************************************************************

\section{First Blue List of the KPNO International Spectroscopic Survey}

\subsection{Selection Criteria}

The selection of ELG candidates from the objective-prism images
is described in detail in KR1.  Here we provide a brief summary of
the key issues.  The selection methods used for the current (blue)
spectral data are identical to those used for the red data in KR1.

The KISS reduction software selects ELG candidates by searching the
extracted objective-prism spectra for objects with 5$\sigma$
emission features.   This is the primary selection criterion of the 
survey.  After this automated selection process, the 
candidate ELGs are checked manually and many spurious sources are 
rejected.  These rejected sources often consist of very faint objects
with noisy spectra, and brighter sources with strong spectral breaks
within the KISS bandpass.  Following the evaluation of the 
software-selected candidates, the spectral images are scanned visually
as a final check for possible 5$\sigma$ sources that were missed
by the software.  These are typically objects with lines very close 
to the red end of the spectra, or bright galaxies with lower equivalent 
width lines where the continuum fitting process overestimates the 
continuum level slightly and hence underestimates the line strength.  
These objects are added to the KISS tables manually.  For the current 
survey list, 17.8\% of the objects were added to the final catalog 
during this visual search.  The combination of the automatic software 
and this visual checking ensures that the sample is quite complete for 
objects with $\ge$5$\sigma$ lines.  

As mentioned in KR1, objects with emission lines between 4$\sigma$ and 
5$\sigma$ are also flagged in the database tables and retained as possible 
ELGs.  This sample of $<$5$\sigma$ sources is not statistically complete, 
and hence is not included in the main survey lists.  Rather, we include 
these additional lower probability sources in a secondary list of ELG 
candidates which should be thought of as a supplement to the main KISS 
catalog.

\subsection{The Survey}

The list of ELG candidates selected in the blue survey is presented in 
Table 2.  Because the survey data includes both photometrically-calibrated 
direct images and spectral images, we are able to 
include a great deal of useful information about each source, such as
accurate photometry and astrometry and estimates of the redshift,
emission-line flux, and equivalent width.  Only the first page of the
table is printed here; the complete table is available in the electronic
version of the paper.

%\placetable{table:tab2}
%\placefigure{fig:find1}
\begin{figure*}[htp]
\vskip -0.5in
\epsfxsize=6.5in
\hskip 0.5in
\epsffile{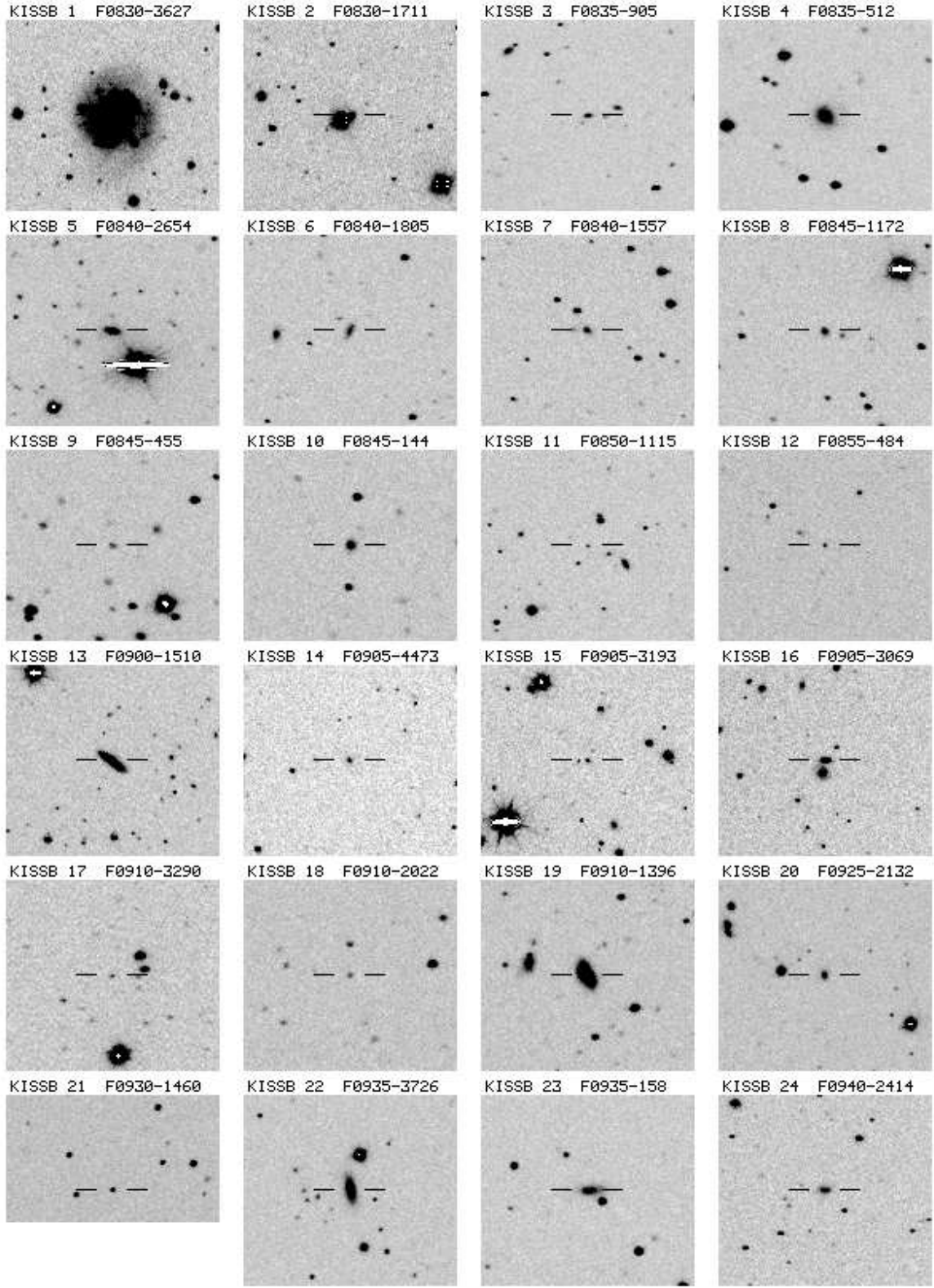}
\figcaption[fig1.eps]{Example of finder charts for the KISS ELG candidates.
Each image is 4.5 $\times$ 4.0 arcmin, with N up, E left.  These finders are 
generated from the direct images obtained as part of the survey.  In all 
cases the ELG candidate is located in the center of the image section 
displayed, and is indicated by the tick marks.\label{fig:find1}}
\end{figure*}

%\placefigure{fig:spec1}
\begin{figure*}[htp]
\vskip -0.5in
\epsfxsize=6.5in
\hskip 0.5in
\epsffile{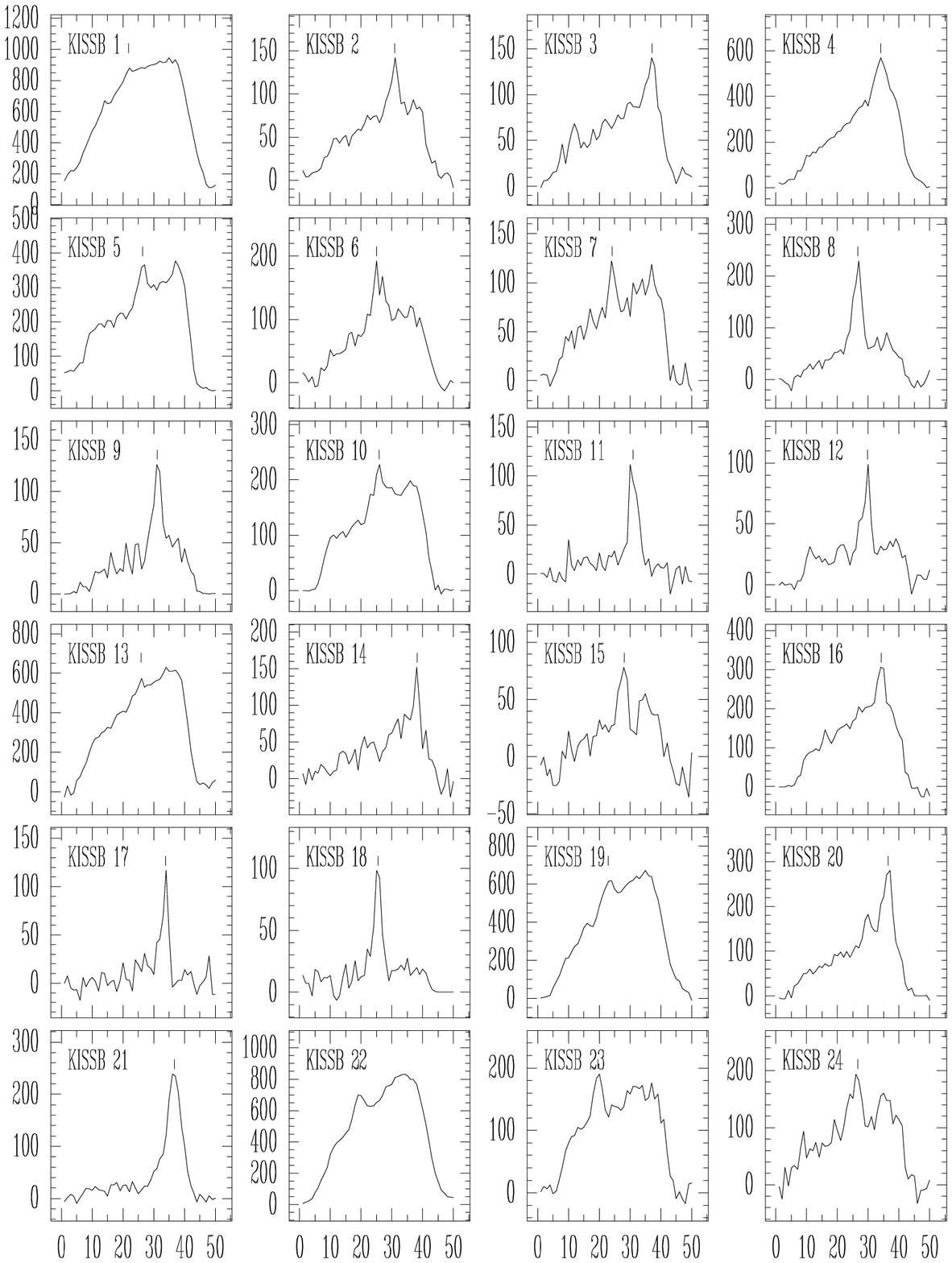}
\figcaption[fig2.eps]{Plots of the objective-prism spectra for the first 24 KISS ELG
candidates listed in Table 1.  The spectral information displayed represents the extracted
spectra present in the KISS database tables.  The location of the putative 
emission line is indicated.\label{fig:spec1}}
\end{figure*}

The contents of the survey table are as follows.  Column 1 gives a running 
number for each object in the survey with the designation KISSB XXXX, where 
KISSB stands for ``KISS blue" survey.  This is to distinguish
it from the red KISS survey (KR1).  Columns 2 and 3 give the object
identification from the KISS database tables, where the first number
indicates the survey field (FXXXX), and the second number is the ID number 
within the table for that galaxy.  This identifier is necessary for locating
the KISS ELGs within the survey database tables.  Columns 4 and 5 list
the right ascension and declination of each object (J2000).  The formal
uncertainties in the coordinates are 0.25 arcsec in right ascension and 0.20 
arcsec in declination.  Column 6 gives the B magnitude, while column 7 lists the 
B$-$V color.  For brighter objects the magnitude estimates have uncertainties
of typically 0.05 magnitude, increasing to $\sim$0.10 magnitude at B = 20.
Paper I includes a complete discussion of the precision of both the astrometry 
and photometry of the KISS objects.  An estimate of the redshift of each
galaxy, based on its objective-prism spectrum, is given in column 8.
This estimate assumes that the emission line seen in the objective-prism 
spectrum is [\ion{O}{3}].  Follow-up spectra for 123 ELG candidates
from the current list show that this assumption is correct in the vast 
majority of cases.  The formal uncertainty in these redshift estimates is 
$\sigma_z$ = 0.0049 (see Section 4.1.3).  There are three objects in the table with
negative redshifts listed.  In all three cases the measured value is just
slightly negative (i.e., within the error quoted above), and two of the
three have redshifts measured from follow-up spectra that are both small but 
positive. Columns 9 and 10 list the emission-line flux (in units of 10$^{-16}$ 
erg/s/cm$^2$) and equivalent width (in \AA) measured from the objective-prism 
spectra.  The calibration of the fluxes is discussed in Section 4.1.2.  These
quantities should be taken as being representative estimates only.
A simple estimate of the reliability of each source, the quality flag (QFLAG),
is given in column 11.  This quantity, assigned during the line measurement 
step of the data processing, is given the value of 1 for high quality 
sources, 2 for lower quality but still reliable objects, and 3 for somewhat 
less reliable sources.  Column 12 lists the KISSR number from KR1 for objects
that were selected in both the red and blue surveys.  The red survey overlaps
the blue for field numbers F1215 and above (excluding fields F1430, F1435, and 
F1440).  There are 125 KISSB ELGs in the overlap area, and 113 (90\%) of these
are also cataloged in either KISSR or KISSRx (see KR1).  Column 13 gives alternate 
identifications for KISS ELGs which have been cataloged previously.  This is 
not an exhaustive cross-referencing, but focuses on previous objective-prism
surveys which overlap part or all of the current survey area: Markarian (1967), 
Case (Pesch \& Sanduleak 1983), Wasilewski (1983), and UCM (Zamorano \etal 1994).  
Also included are objects in common with the {\it Uppsala General Catalogue of 
Galaxies} (UGC, Nilson 1973).

%\placefigure{fig:appmag}
\begin{figure*}[htp]
\vskip -0.4in
\epsfxsize=5.0in
\hskip 1.0in
\epsffile{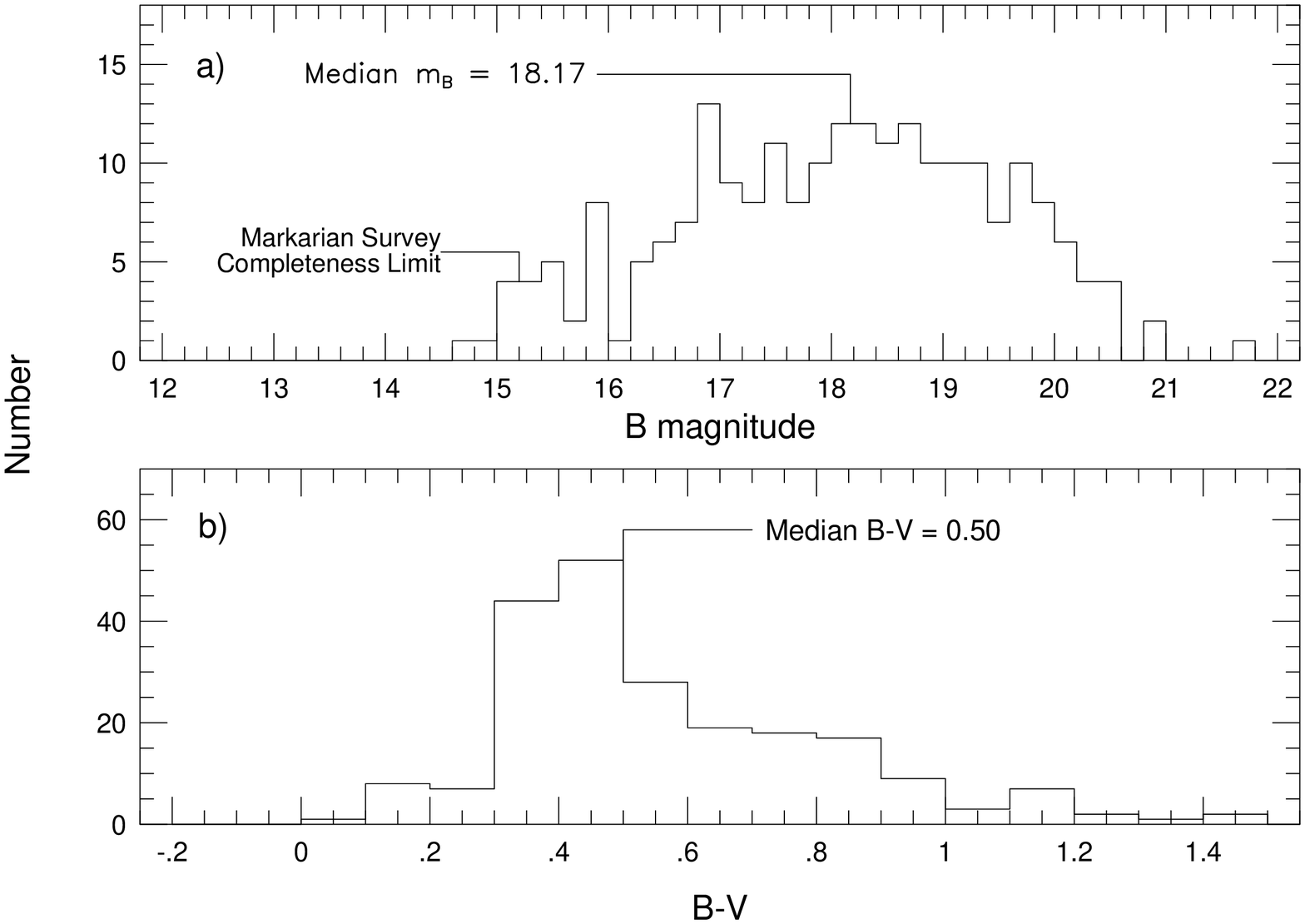}
\figcaption[bmag_col.eps]{(a) Distribution of B-band apparent magnitudes for the
223 ELG candidates in the first [\ion{O}{3}]-selected KISS survey list.  The median 
brightness in the KISS sample is B = 18.17, with 8\% of the galaxies having 
B $>$ 20.  Also plotted, for comparison, is the completeness limit of the 
Markarian survey. (b) Histogram of the B$-$V colors for the 223 ELG candidates.  
The median color of 0.50 is indicated. \label{fig:appmag}}
\end{figure*}

A total of 223 ELG candidates are included in this first list of 
[\ion{O}{3}]-selected KISS galaxies.  The total area covered by the survey 
is 116.6 deg$^2$, meaning that there are 1.91 KISS ELGs per deg$^2$.  This 
compares to the surface density of 0.1 galaxies per deg$^2$ from the Markarian 
survey, and 0.52 per deg$^2$ for the UM survey (MacAlpine \etal 1977), one
of the deepest of the photographic surveys that selected exclusively by line
emission (and primarily by [\ion{O}{3}]).  
Of the total, 119 were assigned quality values of QFLAG = 1 (53.4\%), 89 have 
QFLAG = 2 (39.9\%), and 15 have QFLAG = 3 (6.7\%).  Based on our follow-up spectra
to date, 99\% (71 of 72) of the sources with QFLAG = 1 are {\it bona fide}
emission-line galaxies, compared to 95\% (42 of 44) with QFLAG = 2 and
86\% (6 of 7) with QFLAG = 3.  In total, 97\% of KISSB ELG candidates with 
follow-up spectra are {\it bona fide} emission-line galaxies.
The properties of the KISS galaxy sample are described in the next section.

Figure~\ref{fig:find1} shows an example of the finder charts for the KISS 
ELGs.  These are generated from the direct images obtained as part of the
survey.  Figure~\ref{fig:spec1} displays the extracted spectra derived
from the objective-prism images for the first 24 ELGs in Table 1.  Finder
charts and spectral plots for all 223 objects in the KISS survey are
available in the electronic version of this paper. 

A supplementary table containing an additional 91 ELG candidates is included 
in the appendix of this paper (Table 3).  These galaxies are considered to be lower
probability candidates, having emission lines with strengths between 4$\sigma$
and 5$\sigma$.  These additional galaxies do not constitute a statistically
complete sample, and should therefore be used with caution.  However, there
are likely many interesting objects contained in this supplementary list.
Hence, following the precedent established in KR1, we list them in a separate
table in order to give a full accounting of the ELGs in the fields surveyed.

%************************************************************************

\section{Properties of the KISS ELGs}

\subsection{Observed Properties}

One of the advantages of the KISS observing method is that a large
amount of information can be derived for each ELG candidate from the
survey data themselves.  This reduces the need for follow-up observations.  
The imaging data provide accurate B and V photometry, astrometry, and 
morphological data, while the digital objective-prism spectra allow us 
to measure the position and strength of the observed emission line. 
Of course, the objective-prism spectra are of such low resolution and
small spectral range that one cannot use them to classify the ELGs by 
activity type (e.g., Seyfert vs. starburst).  Therefore, follow-up spectra 
are still necessary in order to develop a more complete understanding of 
the nature of each ELG.  Nonetheless, one can learn a great deal about
the make-up of the KISS ELG sample by examining the properties of the survey
constituents using the survey data alone.  In the following section,
we examine key characteristics of the sample, and compare the properties
of the KISSB galaxies with those from previous objective-prism surveys.

\subsubsection{Magnitude \& Color Distributions}

Figure~\ref{fig:appmag}a displays the B-band apparent magnitude distribution for the KISSB
galaxies, while Figure~\ref{fig:appmag}b plots the B$-$V color distribution.  The median
values of both B and B$-$V for the KISS galaxies are indicated, as is the 
magnitude corresponding to the completeness limit of the Markarian survey 
(Mazzarella \& Balzano 1986), for comparison.  Clearly, the KISS sample probes 
substantially deeper (by $\sim$3 magnitudes) than the Markarian survey.  The 
survey appears to have good sensitivity down to B = 20, beyond which it quickly
drops.  KISS is probably not as sensitive to bright galaxies (B $<$ 15), because the
contrast between the continuum and emission line is not sufficient to allow
us to detect low equivalent width emission.  We can estimate the incompleteness 
at these bright magnitudes in the following way.  Previous studies of [\ion{O}{3}]-selected
ELG samples indicate that approximately 6.7\% of the {\it field-galaxy} population have
strong enough [\ion{O}{3}] emission lines to be detected in a survey like KISSB
(Salzer \etal 1989).  In the area surveyed, there are only 58 CGCG galaxies
(Zwicky \etal 1961) with B $<$ 15, and of these 9 are located in the Coma cluster.
Hence, we predict that there should be 3.3 [\ion{O}{3}]-selected objects in our 
survey area with B $<$ 15, while the actual number found is 2.  Therefore, while we
suspect that we are somewhat less complete at brighter magnitudes, there is no evidence
that KISS is significantly incomplete here.  Certainly KISSB will become progressively 
less sensitive to modest equivalent-width emission lines at brighter magnitudes.

\begin{figure*}[htp]
\vskip -0.4in
\epsfxsize=5.0in
\hskip 1.0in
\epsffile{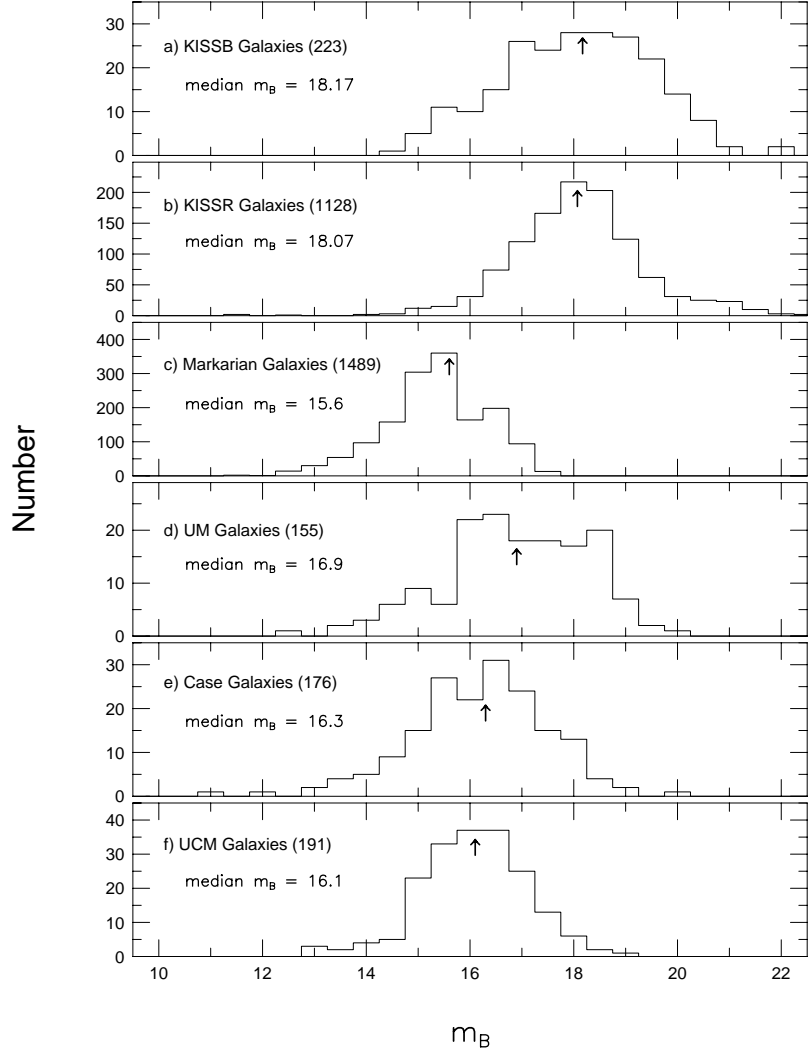}
\figcaption[bmag_comp.eps]{Comparison of the B-band apparent magnitude distributions
for several samples of active galaxies: (a) the KISSB sample from the current paper;
(b) the KISSR H$\alpha$ sample from Salzer \etal (2001); (c) the full Markarian sample,
taken from Mazzarella \& Balzano (1986); (d) the [\ion{O}{3}]-selected UM survey
galaxies (Salzer \etal 1989); (e) the Case Survey galaxies (Salzer \etal 1995); (f)
the H$\alpha$-selected UCM sample (P\'erez-Gonz\'alez \etal 2000).  The number of 
galaxies in each sample is indicated in parentheses next to the survey name.  The median
apparent magnitude is indicated for each sample, and is marked by the arrow in each 
histogram. \label{fig:magcomp}}
\end{figure*}

\begin{figure*}[htp]
\vskip -0.4in
\epsfxsize=5.0in
\hskip 1.0in
\epsffile{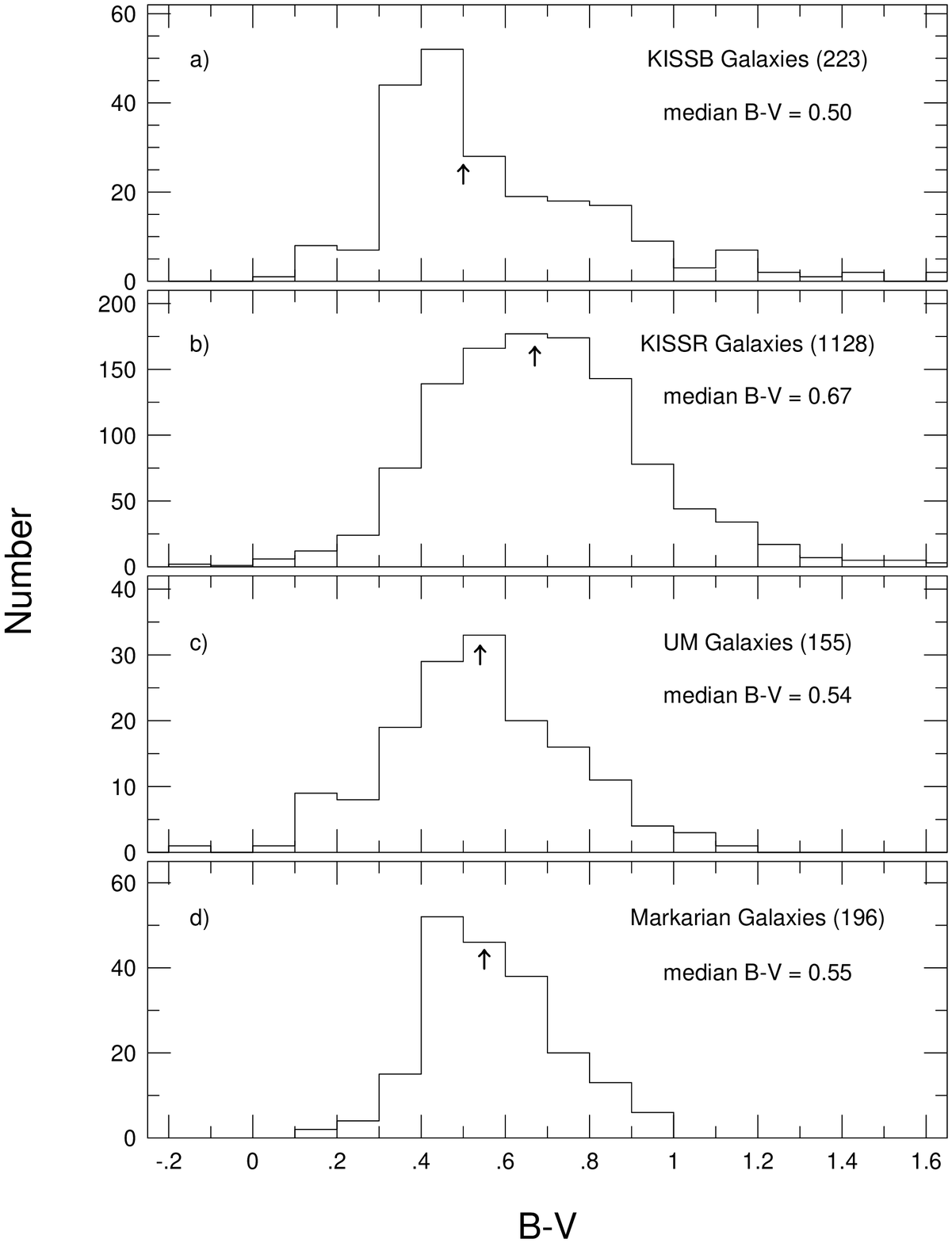}
\figcaption[col_comp.eps]{Comparison of the B$-$V color distributions for (a) the 
KISSB sample from the current paper; (b) the KISSR H$\alpha$ sample from Salzer \etal 
(2001); (c) the [\ion{O}{3}]-selected UM survey galaxies (Salzer \etal 1989); (d) a
large sample of Markarian galaxies with UBV photometry in Huchra (1977).  The 
median color is indicated for each sample. \label{fig:colcomp}}
\end{figure*}

In order to place the magnitude and color distributions for KISSB into
better perspective, we display in Figures~\ref{fig:magcomp} and \ref{fig:colcomp} 
histograms comparing the KISS galaxies to those from other surveys.  
Figure~\ref{fig:magcomp}  plots histograms showing
the distributions of apparent magnitude for KISSB, the first red 
(H$\alpha$-selected) KISS catalog (see KR1), the Markarian survey (Mazzarella \& 
Balzano 1986), the UM survey (Salzer \etal 1989), the Case survey
(Salzer \etal 1995), and the UCM survey (P\'erez-Gonz\'alez \etal 2000).  
The median B magnitudes are indicated for each data set.  Interestingly, the
median depth of KISSB and KISSR are essentially identical: B = 18.17 for the 
blue survey and B = 18.07 for the red. This is true even though the two surveys
are derived from completely independent spectral data.  Despite their similar
depths, the red survey detects nearly ten times more ELGs per
unit area than does KISSB, a testament to the ubiquity of H$\alpha$ emission.  Of
the samples plotted, KISSB compares most closely to the UM survey in terms of
object detection method: both are primarily [\ion{O}{3}]-selected.  KISSB
is roughly 1.3 magnitudes deeper than the UM survey.  Similarly, KISSR and
the UCM survey are both H$\alpha$-selected, with KISSR having a median
brightness a full 2 magnitudes fainter.

%\placefigure{fig:magcomp}

%\placefigure{fig:colcomp}

Figure~\ref{fig:colcomp} shows a similar comparison, but for B$-$V color.  In 
this case, only the KISSR, UM, and Markarian samples have reasonably complete 
BV photometry to allow for this comparison.  Here the B$-$V colors for the Markarian
galaxies are taken from Huchra (1977).  Since the KISSB and UM samples have similar 
selection methods, it is perhaps no surprise that they have similar color distributions.
The UV-excess-selected Markarian sample also exhibits a similar range and median
color.  In contrast, the difference between the blue and red KISS samples
is substantial.  The median colors, B$-$V = 0.50 for KISSB and
B$-$V = 0.67 for KISSR clearly indicate that the different selection methods
result in the detection of a different sample of galaxies.  KISSB is biased
toward detecting galaxies with large [\ion{O}{3}] line strengths.  Such galaxies
tend to be either dwarfish star-forming galaxies, Seyfert galaxies with
strong high-excitation spectra, or starburst galaxies with low amounts
of external extinction.  Hence, blue colors tend to be favored.  The median
B$-$V color is comparable to the mean color for irregular galaxies (Roberts
\& Haynes 1994).  On the other hand, the H$\alpha$-selected red survey is
able to detect {\it both} the intrinsically blue and/or low reddening galaxies,
as well as objects such as starburst nucleus galaxies and LINERs which may have
appreciable H$\alpha$ but weak [\ion{O}{3}].  That is, the red survey appears
to detect the vast majority of the galaxies found by the blue survey in the
areas were they overlap (see Paper I), and in addition finds a large population 
of galaxies that are completely missed in surveys carried out in the blue.
Surveying for galaxies using the H$\alpha$ selection method leads to a larger
sample of ELGs with a different mix of active and star-forming galaxies than
found in the [\ion{O}{3}]-selected KISSB survey.

\subsubsection{Line Strength Distributions}

As mentioned above, the digital nature of the survey data allows us to
obtain estimates of the emission-line strengths directly from the spectral
images.  This is extremely important, as it allows us to quantify the
selection function and completeness limit for the survey directly from
the survey data. Since KISS is a line-selected survey, the depth of the
sample is defined in terms of the emission-line strengths, not the 
apparent magnitudes of the galaxies (Salzer 1989).  With 
previous line-selected objective-prism surveys these issues could only be 
explored using follow-up spectra.  Our method for determining the 
completeness of the KISS survey is detailed in Gronwall \etal 2002a. In 
this section we illustrate the distribution of equivalent widths and line 
fluxes from the blue survey data.

\begin{figure*}[htp]
\vskip -0.4in
\epsfxsize=5.0in
\hskip 1.0in
\epsffile{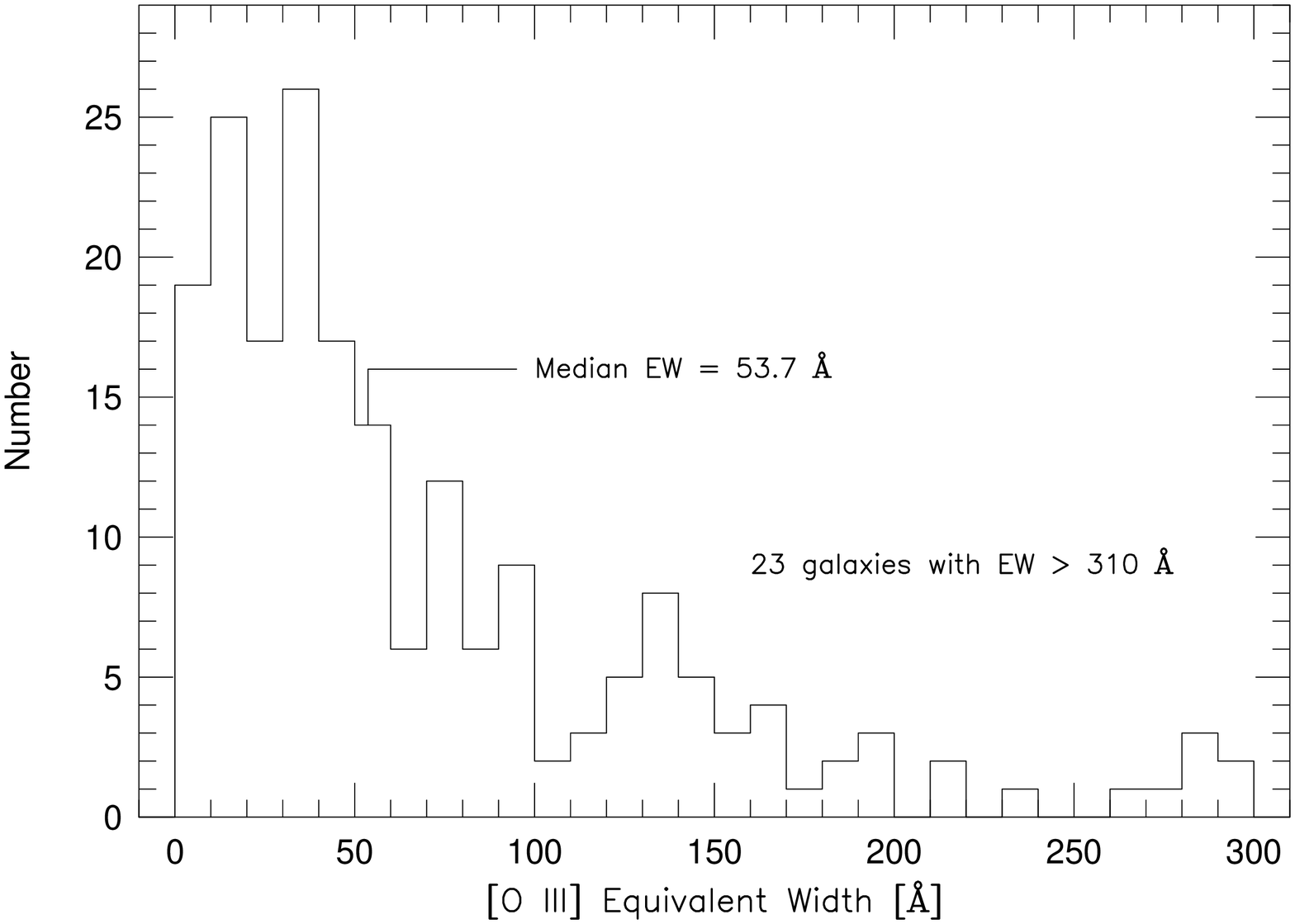}
\figcaption[ew.eps]{Distribution of measured [\ion{O}{3}]
equivalent widths for the KISS ELGs.  The median value of 53.7 \AA\ is 
indicated.  The measurement of equivalent widths from objective-prism 
spectra tends to yield underestimates of the true equivalent widths, so
these values should only be taken as estimates.  The survey appears to 
detect most sources with EW([\ion{O}{3}]) $>$ 30 -- 40 \AA.\label{fig:ew}}
\end{figure*}

Figure~\ref{fig:ew} plots the distribution of [\ion{O}{3}] equivalent widths for 
the 223 KISSB galaxies.  Given the coarse nature of the objective-prism spectra, 
both the equivalent width (EW) and line flux measurements carry with them a
fairly high uncertainty (see KR1 for a complete discussion).  Despite this
caveat, Figure~\ref{fig:ew} is a useful representation of the nature of the KISS ELGs.
The median EW is 54 \AA, somewhat higher than the median for the red survey
(41 \AA).  The distribution of EWs is seen to rise to a peak at 30--40 \AA,
and to contain a large number of galaxies with EW $<$ 30 \AA.  This suggests
that KISS is fairly complete for objects with equivalent widths greater than
$\sim$30 to 40 \AA, but it becomes progressively more incomplete at lower
values.  About 13.5\% of the sample have EWs $>$ 250 \AA, compared to only
3.5\% for the KISSR sample.

%\placefigure{fig:ew}

\begin{figure*}[htp]
\vskip -0.4in
\epsfxsize=5.0in
\hskip 1.0in
\epsffile{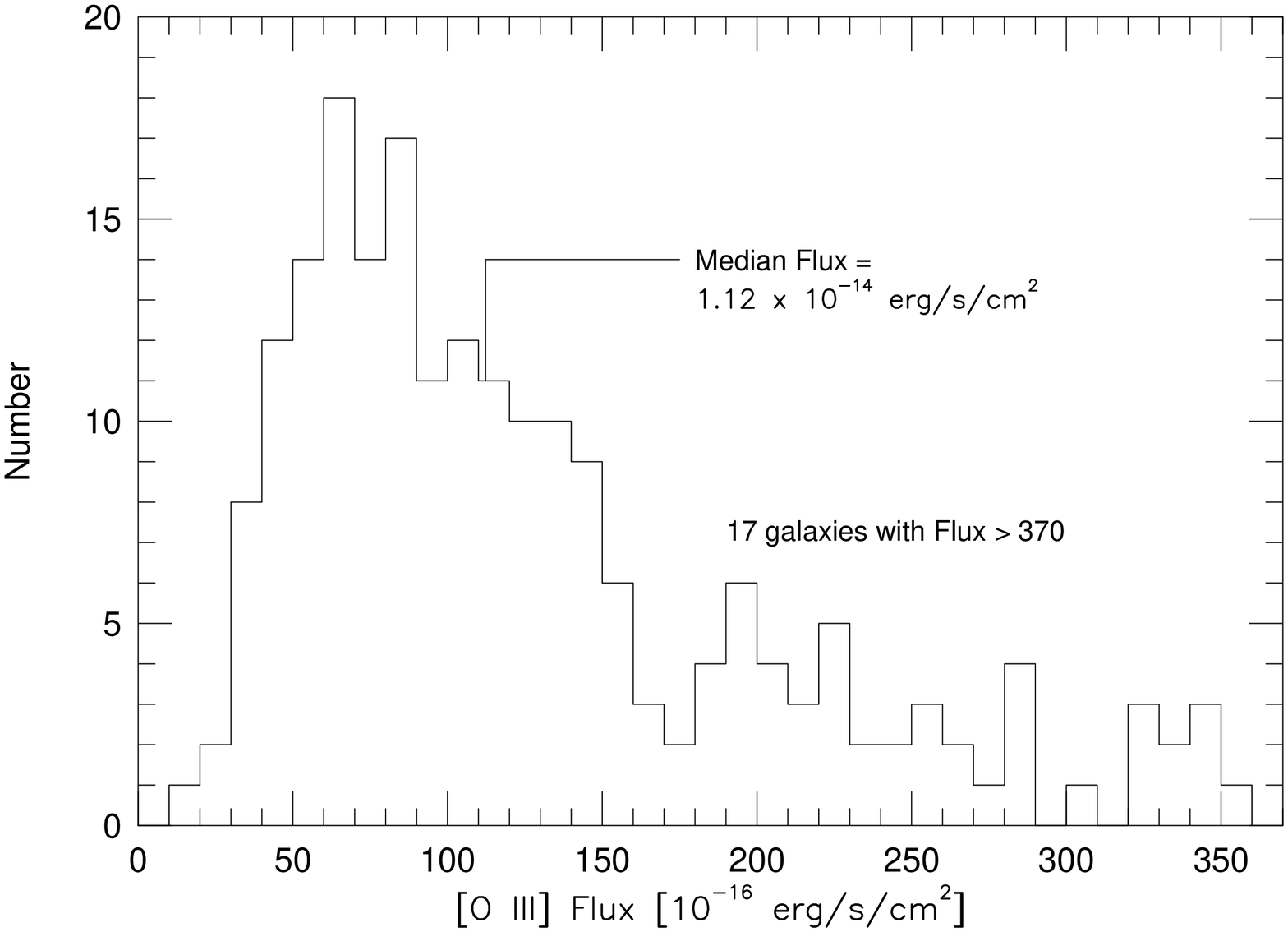}
\figcaption[haflux.eps]{Distribution of [\ion{O}{3}] line fluxes
for the 223 KISS ELGs included in the current survey list.  The median flux
level is indicated.\label{fig:lflux}}
\end{figure*}

The method used to calibrate the objective-prism fluxes is described in
KR1.  In brief, follow-up spectra of non-extended emission-line galaxies
taken under photometric conditions are used to scale the line fluxes
measured from the objective-prism spectra.  The resulting line fluxes 
for the blue survey ELGs are illustrated in Figure~\ref{fig:lflux}.  The 
median flux of 1.12 $\times$ 10$^{-14}$ erg/s/cm$^2$ is about 30\% larger 
than the median H$\alpha$+[\ion{N}{2}] flux for the KISSR sample. The 
faintest objects detected have [\ion{O}{3}] fluxes of $\sim$10$^{-15}$
erg/s/cm$^2$.

%\placefigure{fig:lflux}

\subsubsection{Redshift Distributions}

%\placefigure{fig:zcomp}
\begin{figure*}[htp]
\vskip -0.4in
\epsfxsize=5.0in
\hskip 1.0in
\epsffile{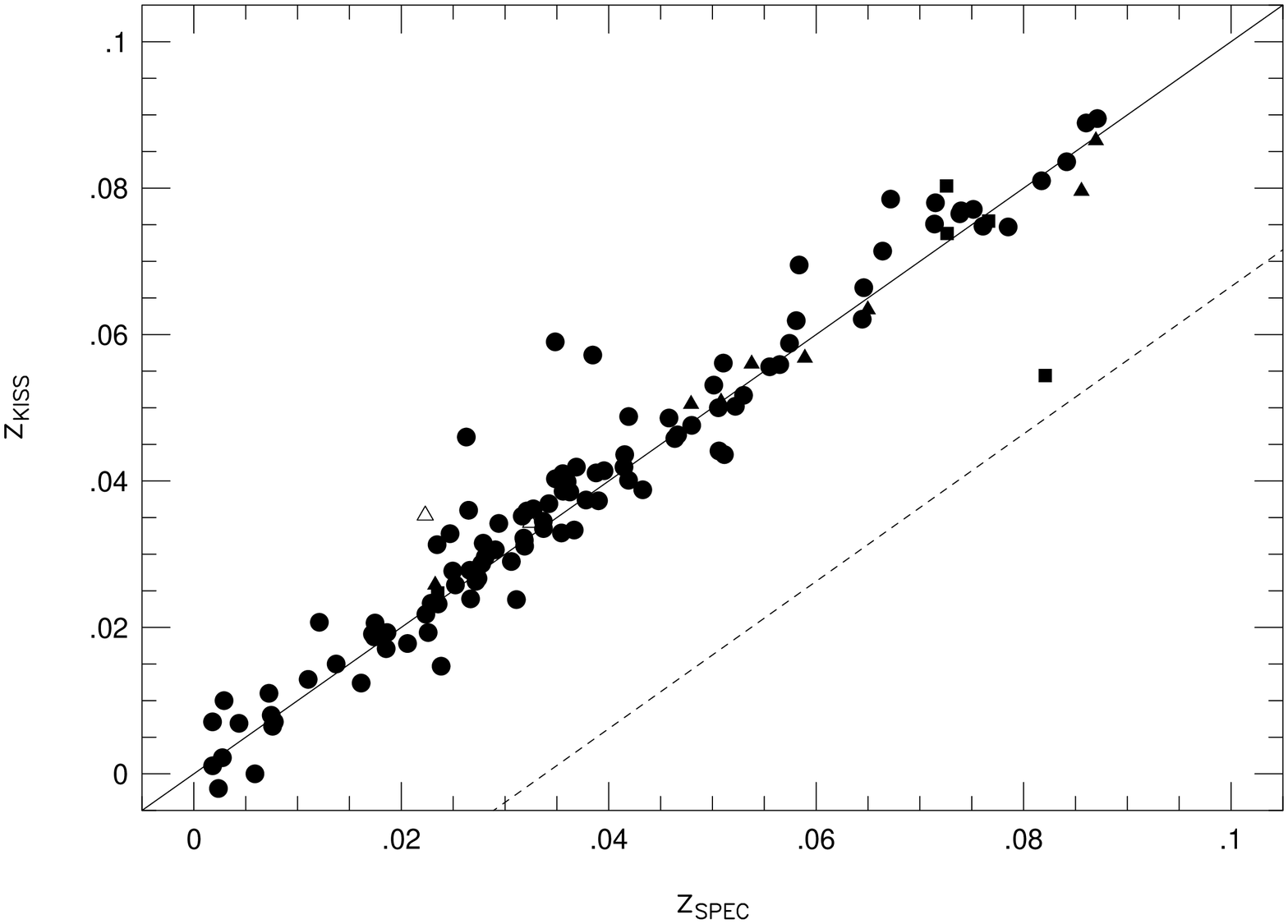}
\figcaption[zcomp.eps]{Comparison between the redshift values estimated from the
objective-prism spectra (z$_{KISS}$) and those obtained from slit spectra (z$_{SPEC}$) 
for 123 KISSB galaxies with follow-up spectra.  Symbols indicate the activity type of
each galaxy: starbursting galaxies are solid circles, Seyfert 1's are solid squares,
Seyfert 2's are solid triangles, and LINERs are open triangles. The solid line indicates 
z$_{KISS}$ = z$_{SPEC}$, while the dashed line shows the expected location of objects 
for which the line detected in the objective-prism spectrum is H$\beta$ rather than 
[\ion{O}{3}]$\lambda$5007.  A single Seyfert 1 galaxy appears to have been 
H$\beta$-selected. \label{fig:zcomp}}
\end{figure*}

Another quantity derived from the objective-prism spectra is the redshift
of each galaxy.  The details of the wavelength calibration method for the
spectral data are given in Paper I and Herrero \etal (2002).  As a check
of the accuracy of the method, we plot in Figure~\ref{fig:zcomp} a comparison between the 
redshifts obtained from follow-up spectroscopy for 123 KISSB ELGs (z$_{SPEC}$)
and the redshift estimates from the survey data (z$_{KISS}$).  The solid
line indicates z$_{KISS}$ = z$_{SPEC}$, while the dashed line shows where objects
for which the line seen in the objective-prism spectrum was H$\beta$ rather than
[\ion{O}{3}]$\lambda$5007 would be located.  One such object appears to have
been found by the survey (a Seyfert 1 galaxy).  There is obviously good agreement
between z$_{KISS}$ and z$_{SPEC}$.
The RMS scatter of the KISS redshifts about the unity line provides an
estimate of the redshift uncertainty associated with z$_{KISS}$.  The
value $\sigma_z$ = 0.0049 is obtained if the single H$\beta$-selected
object is excluded, but the several outlying galaxies above the unity line are
retained.  This corresponds to a velocity uncertainty of 1470 \kms.  
Inspection of the finder charts for the outliers shows that they
are all large galaxies with significant extension in the north-south direction
(the direction of the dispersion in the objective-prism spectra).  In all cases,
the larger than typical redshift offset is most likely due to this extension, 
which can displace the location of the observed emission line relative to the
center of the galaxy.  If these extended outliers in Figure~\ref{fig:zcomp} 
are removed, the value of $\sigma_z$ decreases to 0.0033 (989 \kms),  which 
compares favorably to the value of $\sigma_z$ = 0.0028 found for the KISSR galaxies 
which have follow-up spectra (Paper I).  While this level of precision is not 
adequate for making detailed maps of the galaxian spatial distribution, it is 
sufficient for mapping the coarse structures and for luminosity function work.

A similar comparison between z$_{KISS}$ and z$_{SPEC}$ for the red KISS 
sample revealed a systematic offset in the value of z$_{KISS}$ for redshifts
above about 0.07 (Paper I).  It was concluded that this offset is caused
by the fact that the emission line starts to redshift out of the filter 
bandpass used for the spectral data, and that at higher and higher redshifts
more and more of the red side of the line is missing.  This causes the line
center measured in the objective-prism spectra to be biased to the blue of the
true line center.  No corresponding effect is seen in Figure~\ref{fig:zcomp}, 
although the limited number of follow-up spectra, particularly above z = 0.07, 
make it difficult to assess whether or not a correction for this effect is 
necessary.  Since the blue objective-prism spectra were obtained at somewhat
higher dispersion than the red, it is possible that the onset of the velocity
shift observed in the red survey may be reduced in significance in KISSB.
 
\begin{figure*}[htp]
\vskip -0.4in
\epsfxsize=5.0in
\hskip 1.0in
\epsffile{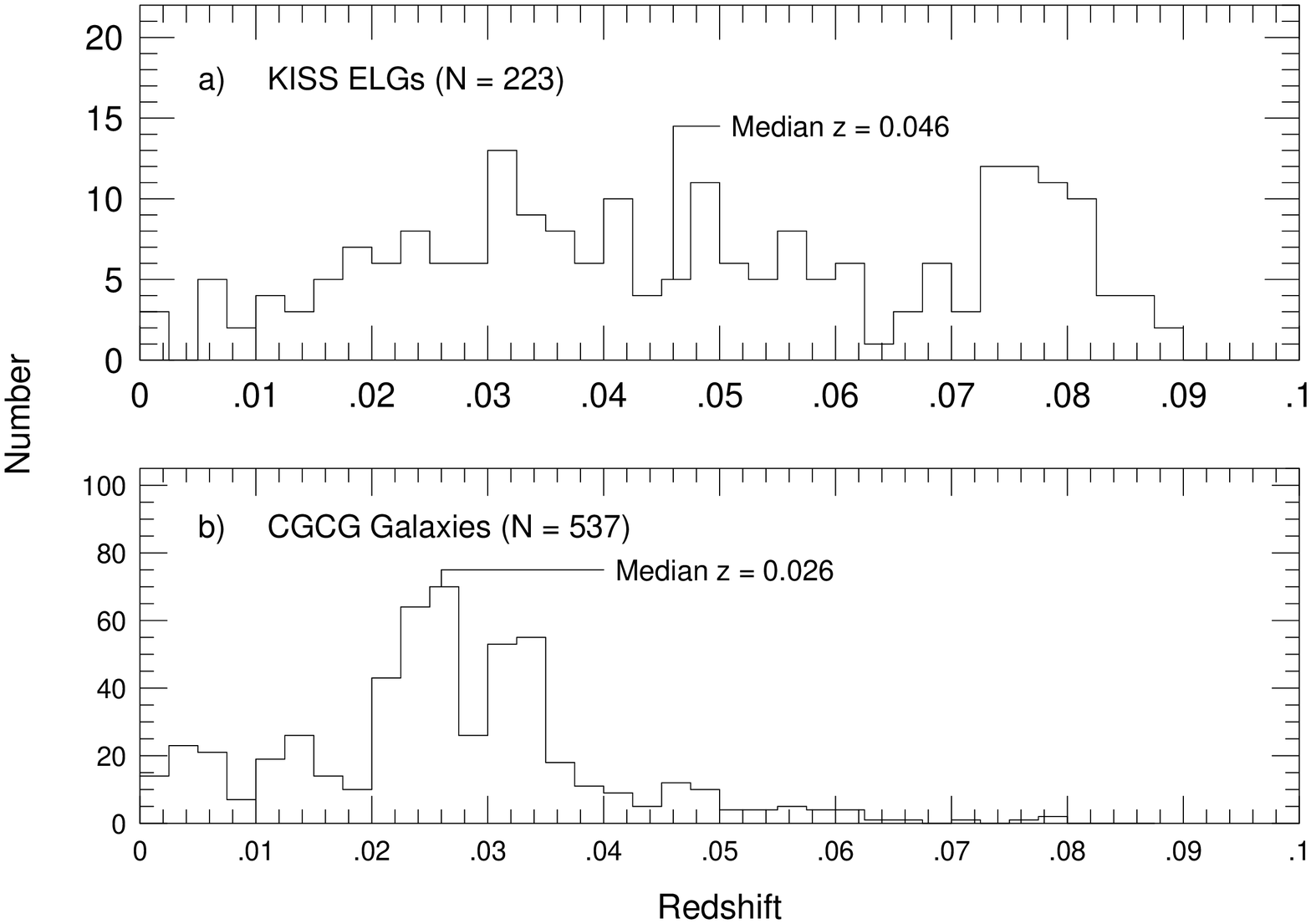}
\figcaption[zhist.eps]{Histograms showing the distribution of redshift for 
(a) the 223 [\ion{O}{3}]-selected KISS ELGs and (b) the 537 ``normal" galaxies 
from the CGCG which are located in the same area of the sky.  The median redshift 
is indicated in both plots.  Note that the number of KISS ELGs remains fairly
constant out to the cut-off of the filter used for the survey. \label{fig:zhist}}
\end{figure*}

Figure~\ref{fig:zhist} illustrates the redshift distributions for (a) the KISSB 
galaxies, and (b) the CGCG (Zwicky \etal 1961) galaxies located in the same area of
the sky.  As described in KR1, this CGCG sample was chosen as a comparison
sample of ``normal" galaxies in the same volume of space covered by the
ELG survey.  To provide adequate numbers, a 3$^\circ$-wide declination strip
of the CGCG is used, compared to the 1$^\circ$-wide KISS strip.  Redshifts
for the CGCG sample come from Falco \etal (1999), and the sample is complete 
for all CGCG galaxies to B = 15.5.  This comparison sample of ``normal"
galaxies is also used in the following section.

%\placefigure{fig:zhist}

The median redshifts for the two samples are indicated in the figure.
The KISS ELGs extend out to much greater distances than do the
magnitude-limited CGCG sample.  In fact, almost no CGCG galaxies are located
at redshifts beyond the median of the KISSB sample.  The numbers of ELGs
remain fairly constant out to z = 0.09, at which point the survey filter
cuts out any higher redshift objects.  The modest peak at z = 0.03 in the 
KISS redshift distribution corresponds roughly to the location of the ``Great 
Wall" seen in the {\it Slice of the Universe} (de Lapparent \etal 1986).

\subsection{Derived Properties}

\subsubsection{Luminosity Distribution}

\begin{figure*}[htp]
\vskip -0.4in
\epsfxsize=5.0in
\hskip 1.0in
\epsffile{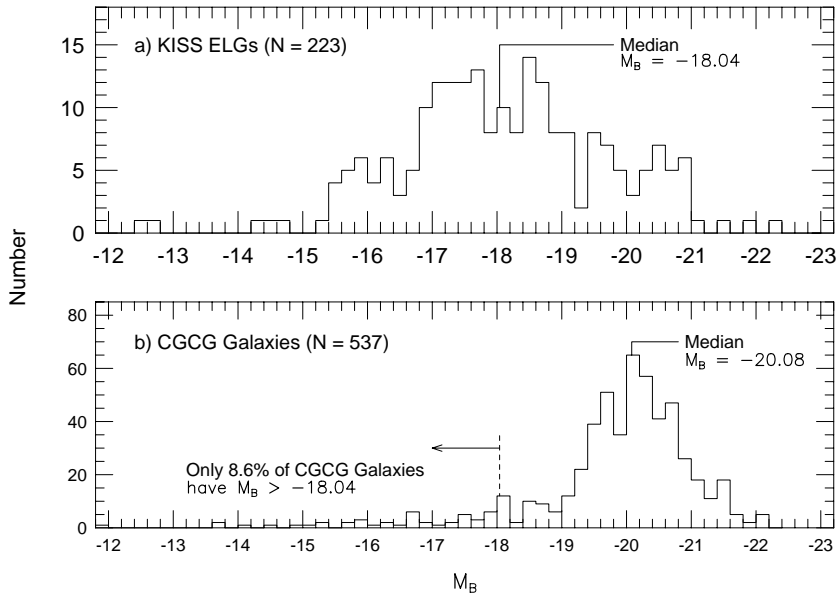}
\figcaption[absmag.eps]{Histograms showing the distribution of blue absolute 
magnitudes for (a) the 223  [\ion{O}{3}]-selected KISS ELGs and (b) the 537
``normal" galaxies from the CGCG that are located in the same area of the sky.  
The median luminosity of each sample is indicated.  The KISS ELG sample is made 
up of predominantly intermediate- and lower-luminosity galaxies, making this 
line-selected sample particularly powerful for studying dwarf galaxies. 
\label{fig:absmag}}
\end{figure*}

We compare the luminosities of the KISSB ELGs with those of the CGCG galaxies
located in the same area of the sky in Figure~\ref{fig:absmag}.  Absolute
magnitudes are computed using the redshifts and apparent magnitudes listed
in Table 2 and assuming a value for the Hubble Constant of H$_o$ = 75 km/s/Mpc.
Corrections for Galactic absorption (A$_B$) have been applied by averaging the values
for all UGC galaxies in each survey field from the compilation of Burstein \&
Heiles (1984).  Since the majority of the survey strip is at high Galactic
latitude, this correction is typically small: 65 of the 102 fields (64\%) have 
A$_B$ $<$ 0.05, and 92 of 102 (90\%) have A$_B$ $<$ 0.10.  The maximum correction
of A$_B$ = 0.26 occurs in the easternmost survey field (F1655).
The median blue absolute magnitude of the KISS ELGs is $-$18.04, which is roughly
two magnitudes fainter than M$^*$, the ``characteristic luminosity" parameter
of the Schechter (1976) luminosity function.  As seen in the lower portion of the
figure, the majority of the CGCG galaxies are more luminous than the KISS ELGs. 
The median absolute magnitude for the comparison sample is $-$20.08, which is
roughly the value of M$^*$.

%\placefigure{fig:absmag}

\begin{figure*}[htp]
\vskip -0.4in
\epsfxsize=5.0in
\hskip 1.0in
\epsffile{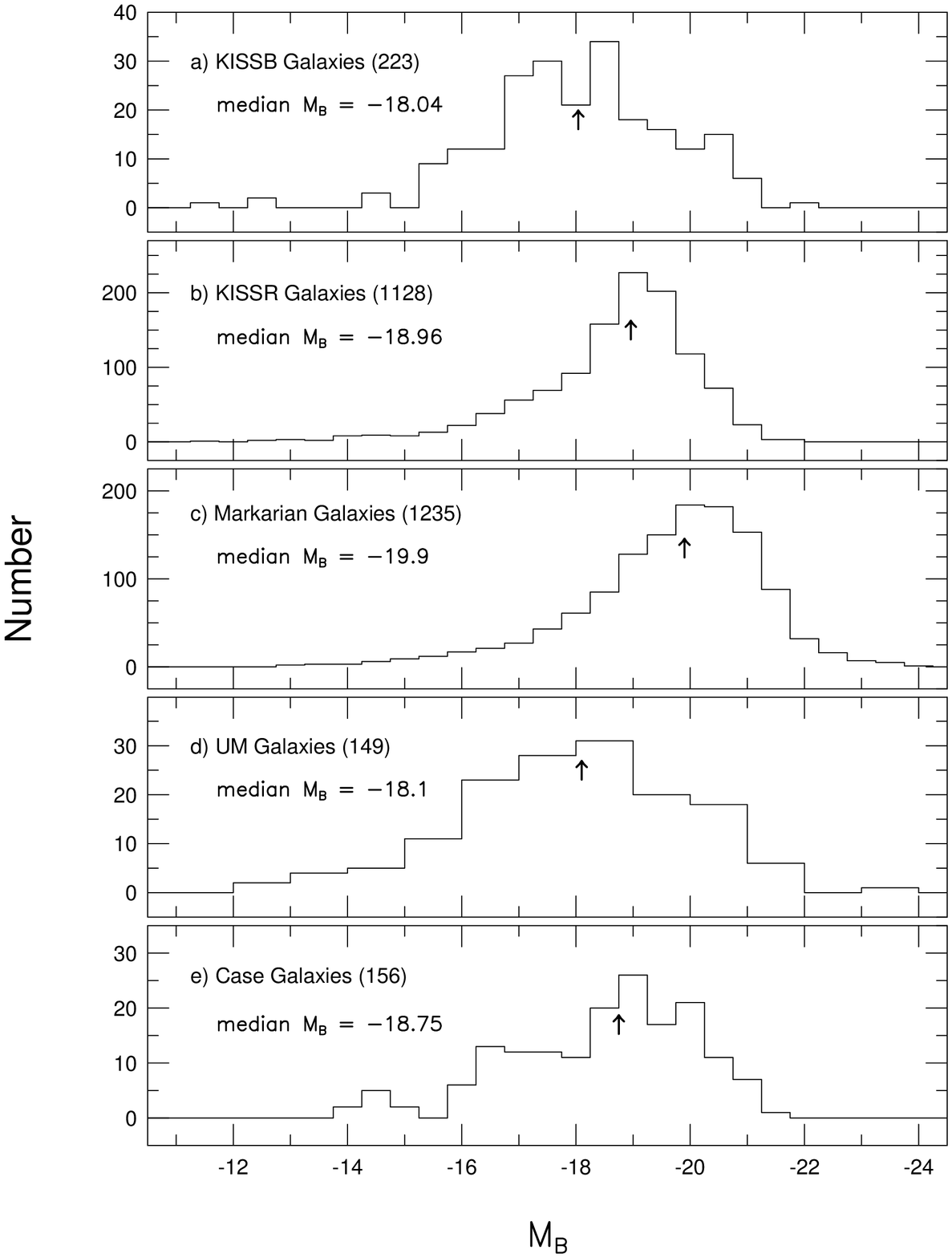}
\figcaption[absmagcomp.eps]{Comparison of the B-band absolute magnitude distributions
for several samples of galaxies: (a) the KISSB sample from the current paper;
(b) the KISSR H$\alpha$ sample from Salzer \etal (2001); (c) the full Markarian sample,
taken from Mazzarella \& Balzano (1986); (d) the [\ion{O}{3}]-selected UM survey
galaxies (Salzer \etal 1989); (e) the Case Survey galaxies (Salzer \etal 1995). 
The median value for each survey is indicated.  The KISSB and UM samples are quite 
comparable, both in terms of the shapes of the distributions and the median values.
The differences between the various surveys can be understood in terms of their
selection methods (see text). \label{fig:abscomp}}
\end{figure*}

In Figure~\ref{fig:abscomp} we present a comparison of the luminosity distributions 
for several other ELG samples.  Included are the absolute magnitude histograms of the
KISSB (this paper), KISSR (KR1), Markarian (Mazzarella \& Balzano 1986), UM (Salzer 
1989), and Case (Salzer \etal 1995) surveys.  The median luminosities are labeled.  
The sample most similar to the KISSB survey are the UM ELGs.  The two have nearly 
identical median values, and similarly shaped distributions.  This should come as no 
surprise, since the two surveys employ the same primary selection method.  The H$\alpha$-selected 
KISSR sample has a median luminosity nearly a full magnitude brighter than KISSB, while 
the UV-excess selected Markarian survey has a luminosity distribution more nearly similar 
to the magnitude-limited CGCG sample plotted in Figure~\ref{fig:absmag}b.  The Case survey 
sample is intermediate between the KISSB and KISSR distributions.

%\placefigure{fig:abscomp}

Figure~\ref{fig:abscomp} illustrates nicely how important the selection method can be 
in defining the make-up of any galaxy sample.  Despite the fact that KISSR and KISSB 
are selected using very similar methods, they contain a substantially different
mix of galaxies.  The only difference is whether [\ion{O}{3}] or H$\alpha$
is the emission line used in the selection of the sample, yet the make-up of
the catalog changes a great deal.  Even greater differences exist between
magnitude-limited samples (the CGCG) and line-selected samples.  Clearly, the
latter are more dwarf dominated than are magnitude-limited samples.  As stressed
in Salzer (1989) and Lee \etal (2000), line-selected surveys like UM and
KISS are excellent for providing large samples of dwarf galaxies at distances
well beyond the Local Supercluster.

\subsubsection{Spatial Distribution}

It is also relevant to examine the spatial distribution of the KISSB galaxies,
and to compare their clustering properties with respect to the CGCG galaxies.
However, due to the coarse nature of the KISS objective-prism redshifts, we limit
ourselves at this time to a simple visual inspection of the spatial distribution 
of the KISS ELGs.  A more comprehensive study will have to wait until higher
quality redshifts have been obtained for the full sample.

\begin{figure*}[htp]
\vskip -0.4in
\epsfxsize=6.5in
\hskip 0.5in
\epsffile{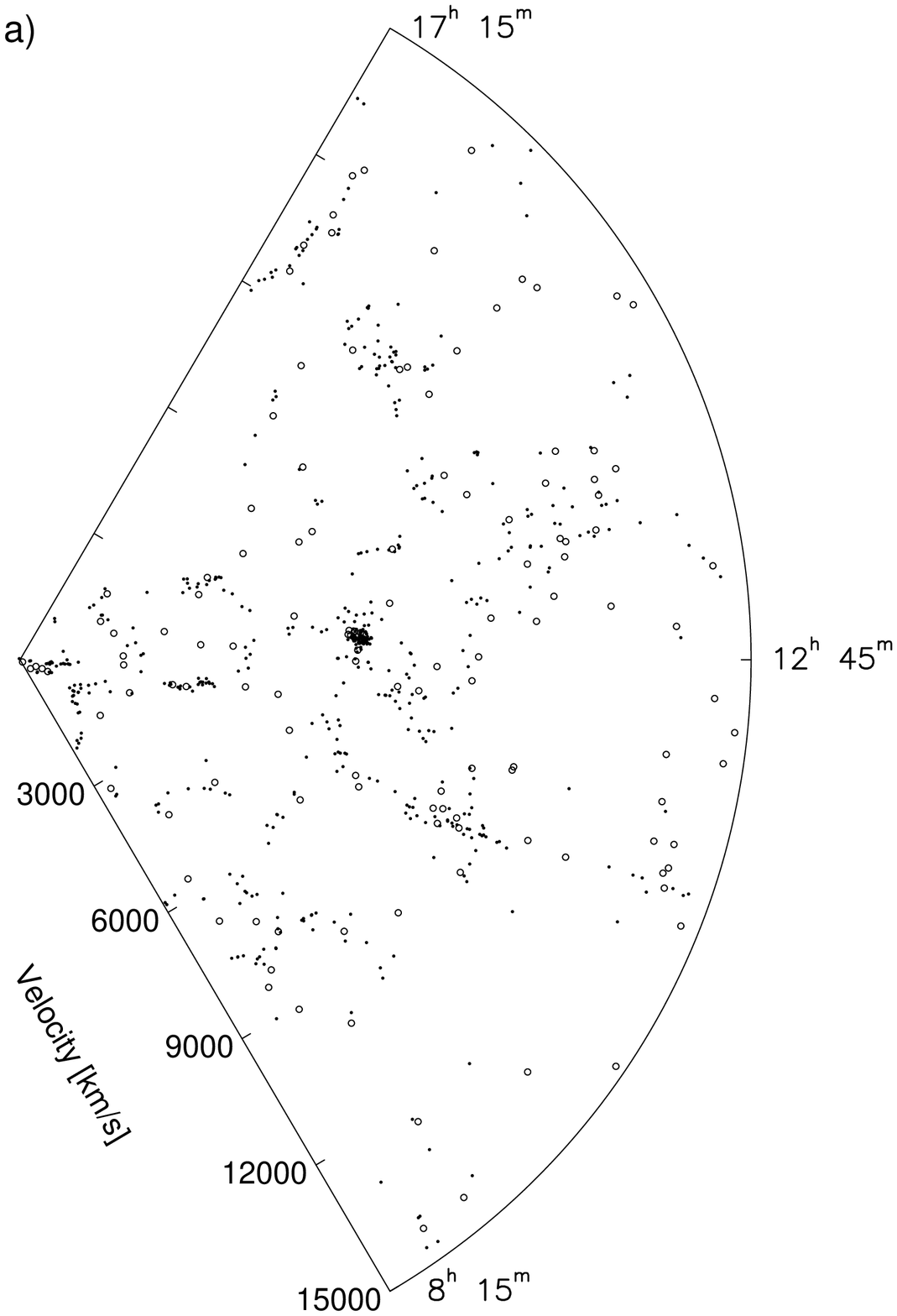}
\figcaption[cone1.eps]{The spatial distribution of the KISS ELGs.  The CGCG
comparison sample is displayed as small dots, while the ELGs are larger, open
symbols.  (a) Velocities plotted out to 15,000 \kms.  The ELGs are seen to 
trace the large-scale structures defined by the CGCG galaxies at low redshift.  
However, they exhibit the appearance of being less tightly clustered.  A large 
number of ELGs are located in voids.  (b) Velocities plotted out to 30,000 
\kms.  At larger distances the ELGs appear to reveal several structures not visible 
in the shallower CfA2 redshift catalog.  The numbers of ELGs remains high out to 
$\sim$25,000 \kms. \label{fig:cone}}
\end{figure*}

Figure~\ref{fig:cone}a shows the spatial distribution of the KISSB galaxies out
to a redshift of 15,000 \kms, where the velocities plotted are based on the redshifts
listed in Table 2.  The ELGs are plotted as the larger open circles.  The CGCG 
galaxy sample, described in the previous section, is plotted for comparison as 
smaller dots.  As was seen
previously in KR1, which covers the R.A. range from 12$^h$ 15$^m$ to 17$^h$ 0$^m$
(i.e., the upper 56\% of the figure), the ELGs tend to lie along the large-scale
structures defined by the CGCG galaxies, but with a tendency toward being somewhat
less clustered.  This behavior has been seen before in several other samples of
ELGs (e.g., Salzer 1989, Rosenberg \etal 1994, Pustil'nik \etal 1995, Popescu 
\etal 1997, Lee \etal 2000).  A number of KISSB galaxies appear to be located 
in voids.  However, {\bf given the low level of precision in the objective-prism
redshifts, the locations of these objects should be considered tentative until 
confirming follow-up spectra can be obtained}.   Once these spectra exist, we
plan to carry out a detailed analysis of the clustering characteristics of the
KISS sample.

%\placefigure{fig:cone}

\begin{figure*}[htp]
\vskip -0.4in
\epsfxsize=6.5in
\hskip 0.5in
\epsffile{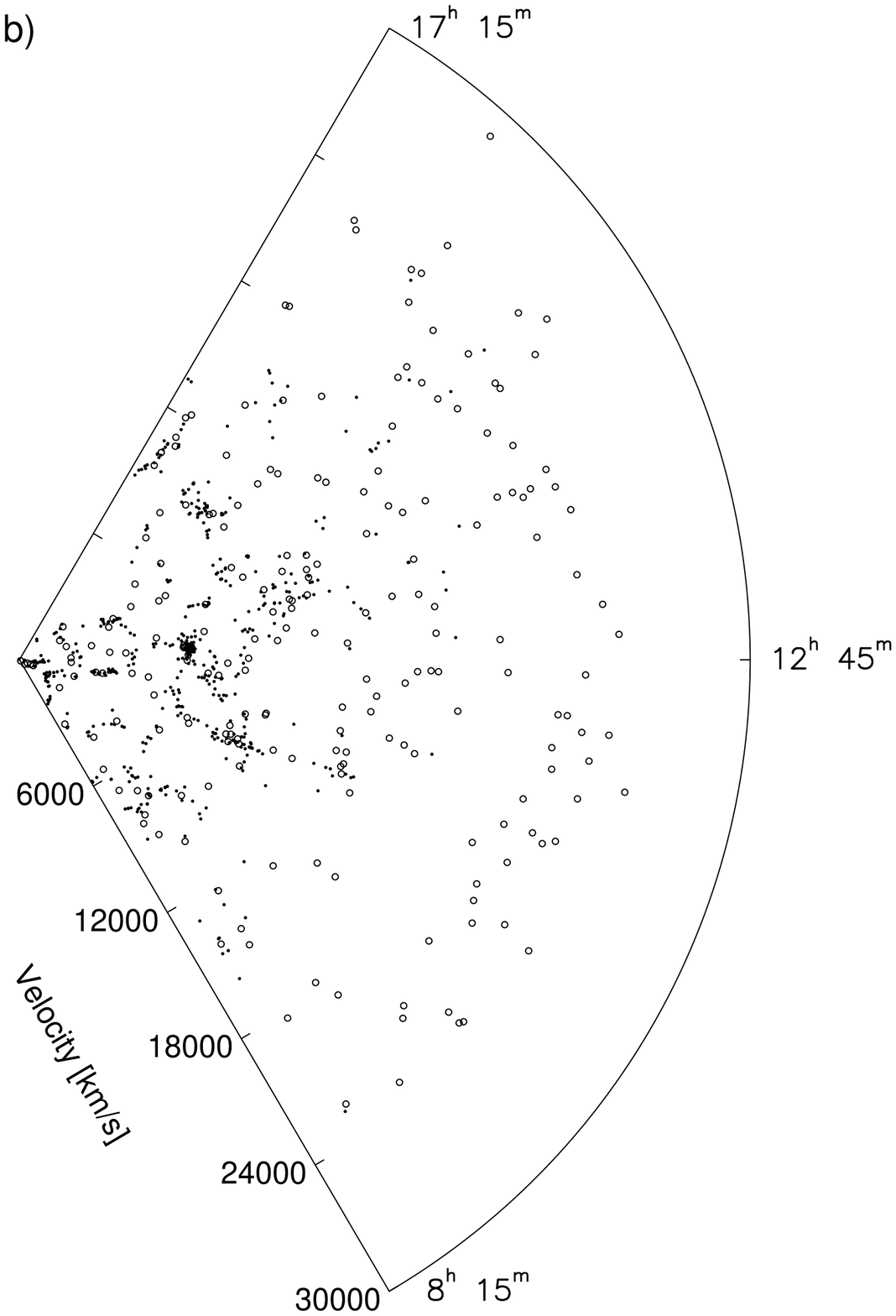}
\end{figure*}

The two samples are plotted again in Figure~\ref{fig:cone}b, this time out to a
redshift limit of 30,000 \kms.  The CGCG sample has few galaxies with redshifts beyond
15,000 \kms, so there is little room for comparisons in this figure.  We note that the
KISSB galaxies populate the region out to $\sim$25,000 \kms\  fairly uniformly, beyond
which point the numbers drop suddenly.  This effect is due to the filter used for the
objective-prism observations, which cuts off the sample quite abruptly between redshifts
of 0.08 and 0.09 (see Figure~\ref{fig:zhist}).  Although the CGCG galaxies do not extend
out to the higher redshift range plotted here, the Century Redshift Survey (CRS, Geller 
\etal 1997; Wegner \etal 2001) sample does.  Comparison with Figure 1 of Geller \etal 
suggests that the higher redshift KISSB galaxies also tend to fall along filaments defined 
by the CRS galaxies, and that the low density region located at $\sim$18,000 \kms\  in the
lower half of the diagram is also present in the CRS data.

A more complete analysis of the spatial distribution and clustering properties of the
KISSB ELGs will be carried out once follow-up spectra for a larger fraction of them 
are available.  Important issues that can be addressed include the environments
of AGN and starburst galaxies as well as the nature of ELGs located in voids.

\subsection{Comparison with Previous Surveys}

The KISS project builds upon the successes of previous objective-prism surveys, some
of which have searched for objects in the same area of the sky.  It is useful,
therefore, to compare the degree to which KISS overlaps with these previous surveys.
It is important to keep in mind the different selection methods of the various surveys
in making such a comparison.  For example, it was found in KR1 that KISS showed excellent
overlap with the previous line-selected surveys that searched the same parts of the sky
(Wasilewski 1983; UCM: Zamorano \etal 1994), but had a lower level of agreement with
color or UV-excess selected samples (Markarian 1967; Case: Pesch \& Sanduleak 1983).

As was already mentioned, there are 125 KISSB objects in the area covered by KR1.  Of
these, 113 (90\%) are cataloged as KISSR or KISSRx objects in KR1.  Since there are
1128 KISSR galaxies and 189 KISSRx objects in this area, less than 10\% of the 
H$\alpha$-selected objects are recovered in the [\ion{O}{3}]-selected sample.  This
is an important point!  The two samples reach to the same emission-line flux levels,
the same apparent magnitudes, and cover the same volumes (due to the survey filters).  
Yet the H$\alpha$-selection process yields $\sim$10 times more objects per unit area than 
does the [\ion{O}{3}]-selection method.  Clearly, strong H$\alpha$ emission is more
ubiquitous than strong [\ion{O}{3}] emission.  Many galaxies which exhibit significant 
activity may have weak lines in the blue portion of the spectrum (e.g., LINERS, highly
reddened starbursts or Seyfert 2s).  For this reason, we have chosen to continue the KISS 
project working exclusively in the red portion of the spectrum for all future ELG catalogs.

Given the results obtained from comparing KISSB to KISSR, it is not surprising that 
KISSB fails to recover a number of objects cataloged in other previous surveys as 
well.  Most of this can be understood as being due to different
selection methods.   There are 15 Markarian galaxies and 114 Case galaxies in the area
covered by the current survey.  Only four Markarian galaxies and 35 Case galaxies are
cataloged in KISSB (27\% and 31\%, respectively).  Since the Markarian survey is a 
UV-excess survey and Case is primarily color selected (Salzer \etal 1995), the low
percentages of overlap with these surveys is no great surprise.  KISSB fares slightly
better with the line-selected UCM (11 of 23, 48\%) and Wasilewski (9 of 11, 82\%) surveys.  
Since UCM is H$\alpha$ selected, one should not expect perfect overlap with KISSB
(even though KISSR recovers 100\% of the UCM galaxies in the areas where the two
surveys overlap).  However, the Wasilewski survey is [\ion{O}{3}] selected, like KISSB,
and also has a much brighter limiting magnitude.   Hence, one might have
expected a higher degree of overlap.  We examined objective-prism data for the two
Wasilewski objects not recovered (Was 12 and Was 88), and found that both are bright 
galaxies with no hint of emission in our spectra.  It seems likely that both possess, at 
best, weak emission lines.  Was 12 is described as a blue elliptical galaxy in Wasilewski (1983),
while Was 88 is listed as having only weak [\ion{O}{3}] emission by Bothun \etal (1989).

%************************************************************************

\section{Summary}

We present our second list of emission-line galaxies discovered as part of the KPNO
International Spectroscopic Survey (KISS).  This is the first list of [\ion{O}{3}]-selected 
ELGs.  A total of 223 galaxies are included in this survey list, which covers an area
of 116.6 deg$^2$.  With a surface density of 1.91 ELGs per deg$^2$, the blue portion
of KISS finds more than 19 times the number of AGNs and starbursting galaxies per unit
area than the Markarian survey, and nearly 4 times the number found by the UM survey,
which used a similar selection method.  An additional 91 ELG candidates detected with a
slightly lower level of statistical significance are cataloged in a supplementary list
(see appendix).  The main advantages of KISS over previous photographic surveys are the 
combination of increased depth and wavelength coverage afforded by the use of a CCD as 
the survey detector.

The digital nature of the survey data, which includes both imaging and spectral observations, 
means that a great deal of information is available for each object in the sample.  In addition 
to tabulating the ELG candidates, Table 2 also includes accurate astrometry, B and V photometry, 
estimates of the redshift, and emission-line strength information for all objects in
the catalog.  This allows us to investigate the properties of the survey constituents
without the need for detailed follow-up observations.  We illustrate the distributions
of apparent magnitudes, colors, emission-line strengths, redshifts, and absolute magnitudes 
for the full sample of KISSB ELGs.  The new catalog of ELGs is found to have an apparent
magnitude distribution very similar to that of KISSR, with a median B magnitude of 18.17.
Not surprisingly, the typical colors of the KISSB galaxies are somewhat bluer than those
for KISSR.  Based on the distribution of measured emission-line equivalent widths, we
estimate that KISSB detects most ELGs in the survey area with [\ion{O}{3}] EW $>$ 30 \AA.
The luminosity distribution is skewed toward lower luminosities (median
M$_B$ = $-$18.04) than KISSR.  This last item is an interesting result, and is due to
the use of [\ion{O}{3}]$\lambda$5007 rather than H$\alpha$ as the primary emission line
used to select the survey constituents.  While KISSR and KISSB are in all other aspects
similar surveys, selecting the ELGs via the two different emission lines changes the
make-up of the respective samples in a major way.

We also compare the properties of the KISS galaxies with those of previous surveys (e.g., 
Markarian, Case, UM, UCM) in order to evaluate the relative strengths of each, as well as 
to better understand how their selection functions help to shape the nature 
of the resulting samples.  The KISSB galaxies are fainter (by 1.3 to 2.6 magnitudes on average)
than the galaxies detected in this representative sample of photographic surveys.  The physical
characteristics of the KISSB ELGS (e.g., color and luminosity) compare most closely with those
of the UM survey, which is also [\ion{O}{3}] selected.   Like the UM survey ELGs, the KISSB
galaxies possess a wide range to physical properties and activity levels, and include both
luminous starbursting and Seyfert galaxies, as well as many intermediate- and low-luminosity
star-forming systems.

We note that, despite the wealth of information available for the KISS ELGs from the survey data 
alone, follow-up spectra are still required in order to confirm each source as a {\it bona fide} 
emission-line galaxy.  In addition, these spectra are necessary for providing more accurate 
redshifts as well as emission-line strengths in order to classify each ELG by their activity 
type (i.e., AGN {\it vs.} starburst).  To date, we have obtained follow-up spectra for slightly
more than half of the KISSB candidates; 119 of 123 KISSB candidates with spectra are found to be
real ELGS (97\%).  Additional follow-up spectroscopic observations are in progress 
for large subsets of the KISS ELG catalogs.

Additional lists of ELG candidates are currently being prepared for publication (e.g., Gronwall
\etal 2002b), and observational data continue to be obtained for new survey areas.  All future
observations will focus on the H$\alpha$ portion of the spectrum, for the reasons described in
the text.  The overall goals of the KISS survey are to cover roughly 300 sq. deg. of sky, and 
to catalog in excess of 5000 ELG candidates.

\acknowledgments

We gratefully acknowledge financial support for the KISS project from an
NSF Presidential Faculty Award to JJS (NSF-AST-9553020), which was instrumental
in allowing for the international collaboration.  Additional support for
this project came from NSF grant AST-9616863 to TXT, and from Kitt Peak National
Observatory, which purchased the special filters used by KISS.  Summer research 
students Michael Santos, Laura Brenneman, and Erin Condy, supported by the Keck 
Northeast Astronomy Consortium student exchange program, helped to reduce the 
survey data presented in the current paper.  We are grateful to Laura Chomiuk,
who assisted in the final production of this paper, and to Katherine Rhode and
Anna Jangren for their critical reading of the manuscript.  Several useful 
suggestions by the anonymous referee helped to improve the presentation of this 
paper.  We thank the numerous colleagues with whom we have 
discussed the KISS project over the past several years, including Jes\'us Gallego, 
Rafael Guzm\'an, Rob Kennicutt, and David Koo.  Finally, we wish to thank the support 
staff of Kitt Peak National Observatory for maintaining the Burrell Schmidt telescope 
and instrument during the early years of the project, and the Astronomy
Department of Case Western Reserve University for taking over this role after 1997.

%************************************************************************

\appendix
\section{Supplementary Table of 4$\sigma$ Objects}

As explained in Section 3, the main selection criterion used to decide whether
or not an object is included in the KISS catalog is the presence of a 5$\sigma$
emission feature in its spectrum.  Because of the high sensitivity of the survey
data, many objects were detected with emission lines that were slightly
weaker than this level.  We made the decision to exclude such objects from the
main survey lists, in order to preserve the statistically complete nature of the
sample.  It was felt that the high degree of reliability of the sample would
be compromised somewhat if these objects were included.  However, rather than 
ignore these weaker-lined ELG candidates entirely, we are publishing them in
a supplementary table.

Listed in Table 3 are 91 ELG candidates that have emission lines detected
at between the 4$\sigma$ and 5$\sigma$ level.  The format of Table 3 is the
same as for Table 2, except that the objects are now labeled with KISSBx
numbers (`x' for extra).  The full version of the table, as well as finder 
charts for all 91 KISSBx galaxies, are available in the electronic version 
of the paper. 

%\placetable{table:tab3}

The characteristics of the supplementary ELG sample are similar to those of the
main survey ELGs, although with some predictable differences.  The median
[\ion{O}{3}] equivalent width is 30.3 \AA, substantially below the value for the
main sample.  The KISSBx galaxies are somewhat fainter (median B magnitude
of 18.76) and significantly redder (median B$-$V = 0.90).  Their median redshift
is significantly higher than that of the main sample (0.069), while their median
luminosity is slightly higher ($-$18.3).  Hence, the supplementary ELG list
appears to be dominated by intermediate luminosity galaxies with a somewhat
lower rate of star-formation activity (lower equivalent widths, redder colors)
than the ELGs in the main sample.

%********************************* REFERENCES***************************

%\clearpage

% Now comes the reference list.  In this document, we used \cite to call
% out citations, so we must use \bibitem in the reference list, which
% means we use the LaTeX thebibliography environment.  Please note that
% \begin{thebibliography} is followed by a null argument.  If you forget
% this, mayhem ensues, and LaTeX will say "Perhaps a missing item?" when
% you run it.  Do not call us, do not send mail when this happens.  Put
% the silly {} after the \begin{thebibliography}.
%
% Each reference has a \bibitem command to define the citation format
% to be placed in the text (in []) and the symbolic tag used for 
% cross referencing (in {}).
%
% See sample1.tex, or the AASTeX guide, for an alternative to the \cite-
% \bibitem command.

%********************************* TABLE 1 *****************************************

\clearpage

\begin{deluxetable}{lccc}
%\scriptsize
%\rotate
\tablenum{1}
\tablecolumns{4}
\tablewidth{0pt}
\tablecaption{KISS Blue Survey Observing Runs\label{table:tab1}}
\tablehead{
\colhead{Dates of Run} & \colhead{Number of} & \colhead{Number of} & \colhead{Number of}\\
& \colhead{Nights\tablenotemark{a}} & \colhead{Fields -- Direct\tablenotemark{b}} & \colhead{Fields -- Spectral\tablenotemark{b}}\\
\colhead{(1)}&\colhead{(2)}&\colhead{(3)}&\colhead{(4)}
}
\startdata
March 19 -- 23, 1994\tablenotemark{c} & 2 & ... & ... \\
April 13 -- 17, 1994 & 4 & ... & 19 \\
April 18 -- 24, 1995 & 5 & 5 & 11 \\
April 29 -- May 4, 1995 & 5 & 16 & 13 \\
March 12 -- 18, 1996 & 4 & 10 & 17 \\
May 8 -- 23, 1996 & 15 & 54 & 32 \\
February 12 -- 15, 1997 & 2 & 12 & ... \\
April 30 -- May 2, 1997 & 3 & 5 & 10 \\

\enddata
\tablenotetext{a}{Number of nights during run that data were obtained.}
\tablenotetext{b}{Number of survey fields observed.}
\tablenotetext{c}{Initial run -- only test data obtained.}
\end{deluxetable}

%********************************* TABLE 2 *****************************************

\clearpage

\renewcommand{\arraystretch}{.6}

\begin{deluxetable}{rrrrrrrrrrcrl}
\tabletypesize{\scriptsize}
%\rotate
\tablecolumns{13}
\tablewidth{0pt}
\tablecaption{List of Candidate ELGs\tablenotemark{1}\label{table:tab2}}
\tablenum{2}
\tablehead{
\\
\colhead{KISSB}&\colhead{Field}&\colhead{ID}&\colhead{R.A.}
&\colhead{Dec.}&\colhead{B}
&\colhead{B$-$V}&\colhead{z$_{KISS}$}&\colhead{Flux\tablenotemark{a}}
&\colhead{EW}&\colhead{Qual.}&\colhead{KISSR}&\colhead{Comments}\\
\colhead{\#}&&&\colhead{(J2000)}&\colhead{(J2000)}&&&&&\colhead{[\AA]}&&\\
\colhead{(1)}&\colhead{(2)}&\colhead{(3)}&\colhead{(4)}
&\colhead{(5)}&\colhead{(6)}&\colhead{(7)}
&\colhead{(8)}&\colhead{(9)}&\colhead{(10)}&\colhead{(11)}
&\colhead{(12)}&\colhead{(13)}
}
\startdata
\\
    1&F0830 & 3627& 8 33 23.1&29 32 18.5&  15.38&   0.46& 0.0110&   36&    1& 1&\phm{imav}     & UGC 4469                      \\
    2&F0830 & 1711& 8 34 35.6&29 17 00.2&  18.18&   0.20& 0.0491&   86&   35& 2&\phm{imav}     &\phm{imaveryveryverylongstring}\\
    3&F0835 &  905& 8 39 49.3&28 57 48.2&  18.20&   0.67& 0.0787&   80&   33& 1&\phm{imav}     &\phm{imaveryveryverylongstring}\\
    4&F0835 &  512& 8 40 02.4&29 49 02.6&  15.99&   0.98& 0.0634&  345&   29& 1&\phm{imav}     &\phm{imaveryveryverylongstring}\\
    5&F0840 & 2654& 8 42 45.4&29 23 59.6&  16.28&   0.96& 0.0288&  164&   16& 1&\phm{imav}     &\phm{imaveryveryverylongstring}\\
    6&F0840 & 1805& 8 43 37.7&29 19 21.8&  17.14&   0.27& 0.0230&  153&   44& 1&\phm{imav}     &\phm{imaveryveryverylongstring}\\
    7&F0840 & 1557& 8 43 52.3&29 26 14.6&  17.62&   0.56& 0.0193&   83&   32& 2&\phm{imav}     &\phm{imaveryveryverylongstring}\\
    8&F0845 & 1172& 8 49 01.3&29 29 16.6&  17.55&   0.39& 0.0313&  418&  195& 1&\phm{imav}     &\phm{imaveryveryverylongstring}\\
    9&F0845 &  455& 8 49 59.1&29 40 51.1&  18.64&   0.46& 0.0516&  208&  138& 2&\phm{imav}     &\phm{imaveryveryverylongstring}\\
   10&F0845 &  144& 8 50 25.0&29 40 51.4&  16.96&   0.70& 0.0279&   79&   11& 1&\phm{imav}     &\phm{imaveryveryverylongstring}\\
 \\
   11&F0850 & 1115& 8 54 05.2&29 07 55.8&  19.74&   0.57& 0.0491&  198&  456& 2&\phm{imav}     &\phm{imaveryveryverylongstring}\\
   12&F0855 &  484& 9 00 08.5&28 55 24.2&  19.13&   0.38& 0.0428&  100&  132& 1&\phm{imav}     &\phm{imaveryveryverylongstring}\\
   13&F0900 & 1510& 9 04 09.5&29 48 20.4&  15.41&   0.45& 0.0246&   35&    2& 2&\phm{imav}     &\phm{imaveryveryverylongstring}\\
   14&F0905 & 4473& 9 05 41.3&28 56 36.5&  19.14&   1.26& 0.0813&   71&   62& 2&\phm{imav}     &\phm{imaveryveryverylongstring}\\
   15&F0905 & 3193& 9 07 07.2&29 06 05.5&  18.72&   0.14& 0.0344&   69&  126& 2&\phm{imav}     &\phm{imaveryveryverylongstring}\\
   16&F0905 & 3069& 9 07 08.5&29 28 49.7&  16.83&   0.50& 0.0621&  118&   26& 1&\phm{imav}     &\phm{imaveryveryverylongstring}\\
   17&F0910 & 3290& 9 11 13.5&29 46 22.1&  19.63&   0.44& 0.0656&  171&  510& 2&\phm{imav}     &\phm{imaveryveryverylongstring}\\
   18&F0910 & 2022& 9 13 05.0&28 48 44.4&  19.54&   0.56& 0.0262&  190&  338& 1&\phm{imav}     &\phm{imaveryveryverylongstring}\\
   19&F0910 & 1396& 9 13 48.7&29 10 21.8&  15.03&   0.52& 0.0150&  280&   14& 1&\phm{imav}     & UGC 4860, CG 11               \\
   20&F0925 & 2132& 9 27 25.4&29 25 18.6&  17.26&   0.36& 0.0748&  221&   56& 1&\phm{imav}     &\phm{imaveryveryverylongstring}\\
 \\
   21&F0930 & 1460& 9 33 37.3&28 45 32.7&  19.63&   1.69& 0.0737&  841& 2870& 1&\phm{imav}     &\phm{imaveryveryverylongstring}\\
   22&F0935 & 3726& 9 36 07.6&29 06 44.7&  15.32&   0.52&-0.0004&  336&   19& 1&\phm{imav}     & CG 23                         \\
   23&F0935 &  158& 9 40 12.7&29 35 29.8&  16.32&   0.24& 0.0012&  143&   37& 1&\phm{imav}     &\phm{imaveryveryverylongstring}\\
   24&F0940 & 2414& 9 42 40.6&28 52 00.4&  17.64&   0.35& 0.0296&   70&   42& 2&\phm{imav}     &\phm{imaveryveryverylongstring}\\
   25&F0940 & 1009& 9 44 18.7&29 27 47.2&  19.60&   1.83& 0.0793&  126&   72& 1&\phm{imav}     &\phm{imaveryveryverylongstring}\\
   26&F0940 &  776& 9 44 44.3&28 46 34.4&  18.24&   0.56& 0.0816&   51&   24& 3&\phm{imav}     &\phm{imaveryveryverylongstring}\\
   27&F0940 &  261& 9 45 12.7&29 42 54.9&  20.04&   0.65& 0.0460&  159&  464& 2&\phm{imav}     &\phm{imaveryveryverylongstring}\\
   28&F0940 &  146& 9 45 23.1&29 23 21.9&  16.92&   0.76& 0.0820&  100&   19& 2&\phm{imav}     &\phm{imaveryveryverylongstring}\\
   29&F1005 & 3849&10 06 18.1&28 56 40.8&  14.64&   0.52& 0.0069&  482&   21& 1&\phm{imav}     & CG 50                         \\
   30&F1005 & 3288&10 06 48.2&29 29 10.2&  16.89&   0.65& 0.0707&   59&    7& 2&\phm{imav}     &\phm{imaveryveryverylongstring}\\
 \\

\enddata

\tablenotetext{1}{Note.--- The complete version of this table is presented in the electronic 
edition of the Journal.  A portion is shown here for guidance regarding
its content and format.}
\tablenotetext{a}{Units of 10$^{-16}$ erg/s/cm$^2$}
%\tablenotetext{b}{Test}
\end{deluxetable}

%********************************* TABLE 3 *****************************************

\clearpage

\renewcommand{\arraystretch}{.6}

\begin{deluxetable}{rrrrrrrrrrcrl}
\tabletypesize{\scriptsize}
\tablecolumns{13}
\tablewidth{0pt}
\tablecaption{List of 4$\sigma$ Candidate ELGs\tablenotemark{1}\label{table:tab3}}
\tablenum{3}
\tablehead{
\\
\colhead{KISSBx}&\colhead{Field}&\colhead{ID}&\colhead{R.A.}
&\colhead{Dec.}&\colhead{B}
&\colhead{B$-$V}&\colhead{z$_{KISS}$}&\colhead{Flux\tablenotemark{a}}
&\colhead{EW}&\colhead{Qual.}&\colhead{KISSR}&\colhead{Comments}\\
\colhead{\#}&&&\colhead{(J2000)}&\colhead{(J2000)}&&&&&\colhead{[\AA]}&&\\
\colhead{(1)}&\colhead{(2)}&\colhead{(3)}&\colhead{(4)}
&\colhead{(5)}&\colhead{(6)}&\colhead{(7)}
&\colhead{(8)}&\colhead{(9)}&\colhead{(10)}&\colhead{(11)}
&\colhead{(12)}&\colhead{(13)}
}
\startdata
\\
    1&F0845 & 1865& 8 48 10.5&29 29 28.9&  15.61&   0.89& 0.0806&   29&    1& 2&
\phm{imav}     &\phm{imaveryveryverylongstring}\\
    2&F0850 & 3556& 8 51 18.4&29 23 59.0&  17.91&   0.92& 0.0851&   51&   17& 2&
\phm{imav}     &\phm{imaveryveryverylongstring}\\
    3&F0850 &   47& 8 55 18.8&29 51 09.3&  19.21&   1.68& 0.0838&   57&   25& 2&
\phm{imav}     &\phm{imaveryveryverylongstring}\\
    4&F0855 &  458& 9 00 10.8&28 55 41.8&  19.27&   0.30& 0.0432&   33&   30& 2&
\phm{imav}     &\phm{imaveryveryverylongstring}\\
    5&F0900 &  630& 9 04 58.8&29 51 36.9&  17.51&   0.50& 0.0798&   21&    5& 3&
\phm{imav}     &\phm{imaveryveryverylongstring}\\
    6&F0910 & 1782& 9 13 17.1&29 07 00.0&  16.67&   0.48& 0.0782&   96&   20& 1&
\phm{imav}     &\phm{imaveryveryverylongstring}\\
    7&F0910 & 1363& 9 13 53.2&29 04 41.0&  16.71&   0.77& 0.0560&   91&   11& 1&
\phm{imav}     &\phm{imaveryveryverylongstring}\\
    8&F0915 & 2738& 9 18 32.0&29 43 37.1&  18.66&   1.69&-0.0119&  157&   87& 2&
\phm{imav}     &\phm{imaveryveryverylongstring}\\
    9&F0935 & 2336& 9 37 41.5&29 15 19.3&  19.78&   1.51& 0.0296&   32&   51& 2&
\phm{imav}     &\phm{imaveryveryverylongstring}\\
   10&F0940 & 4063& 9 40 31.8&28 58 09.8&  15.83&   0.46& 0.0399&   24&    1& 2&
\phm{imav}     &UGC 5154, CG 27                \\
 \\
   11&F0940 & 2256& 9 42 51.7&28 59 60.0&  16.80&   0.32& 0.0331&   29&    5& 2&
\phm{imav}     &\phm{imaveryveryverylongstring}\\
   12&F0940 &  279& 9 45 12.3&29 39 32.6&  19.34&   1.08& 0.0548&   56&   63& 3&
\phm{imav}     &\phm{imaveryveryverylongstring}\\
   13&F0950 & 2455& 9 51 05.6&29 47 29.7&  19.46&   1.00& 0.0716&   78&  133& 2&
\phm{imav}     &\phm{imaveryveryverylongstring}\\
   14&F1000 &  887&10 04 28.1&29 14 42.3&  18.99&   0.43& 0.0443&   93&  112& 2&
\phm{imav}     &\phm{imaveryveryverylongstring}\\
   15&F1000 &  762&10 04 38.6&29 11 46.7&  18.65&   0.41& 0.0164&   93&  109& 2&
\phm{imav}     &\phm{imaveryveryverylongstring}\\
   16&F1005 & 3064&10 07 00.4&29 43 17.3&  19.42&   1.41& 0.0363&   37&   34& 2&
\phm{imav}     &\phm{imaveryveryverylongstring}\\
   17&F1005 &  511&10 09 45.1&29 39 32.1&  19.05&   1.26& 0.0765&   41&   27& 2&
\phm{imav}     &\phm{imaveryveryverylongstring}\\
   18&F1010 & 4499&10 10 25.5&28 53 54.7&  17.21&   0.53& 0.0417&   47&    8& 2&
\phm{imav}     &\phm{imaveryveryverylongstring}\\
   19&F1015 &  731&10 19 24.9&29 45 03.7&  17.02&   0.95& 0.0499&   57&   10& 2&
\phm{imav}     &\phm{imaveryveryverylongstring}\\
   20&F1035 & 1675&10 38 19.5&29 26 23.8&  20.78&   0.90& 0.0861&   60&  348& 2&
\phm{imav}     &\phm{imaveryveryverylongstring}\\
 \\
   21&F1045 & 3716&10 45 20.1&29 29 09.6&  17.56&   0.99& 0.0551&   20&    3& 3&
\phm{imav}     &\phm{imaveryveryverylongstring}\\
   22&F1050 & 1288&10 53 37.3&29 21 34.5&  19.21&   1.43& 0.0667&   58&   30& 2&
\phm{imav}     &\phm{imaveryveryverylongstring}\\
   23&F1055 & 2916&10 56 26.6&28 51 02.4&  18.83&   0.61& 0.0712&   73&   96& 3&
\phm{imav}     &\phm{imaveryveryverylongstring}\\
   24&F1100 & 3477&11 02 11.2&29 06 29.6&  19.20&   0.76& 0.0672&   50&   66& 2&
\phm{imav}     &\phm{imaveryveryverylongstring}\\
   25&F1110 & 3530&11 10 21.2&29 28 05.3&  18.24&   0.10& 0.0750&   45&   35& 2&
\phm{imav}     &\phm{imaveryveryverylongstring}\\
   26&F1115 & 1748&11 18 05.7&29 23 54.9&  18.62&   0.27& 0.0488&   39&   28& 2&
\phm{imav}     &\phm{imaveryveryverylongstring}\\
   27&F1120 & 3566&11 20 30.5&29 12 02.9&  16.99&   1.11& 0.0282&   58&    9& 1&
\phm{imav}     &\phm{imaveryveryverylongstring}\\
   28&F1140 & 3025&11 40 59.6&29 13 30.3&  20.08&   0.49& 0.0878&   51&   92& 2&
\phm{imav}     &\phm{imaveryveryverylongstring}\\
   29&F1140 & 1881&11 42 43.8&28 57 20.3&  18.98&   0.36& 0.0591&   59&   52& 2&
\phm{imav}     &\phm{imaveryveryverylongstring}\\
   30&F1145 & 3182&11 46 01.3&29 24 52.5&  19.57&   1.34& 0.0819&  120&  339& 2&
\phm{imav}     &\phm{imaveryveryverylongstring}\\

\enddata
\tablenotetext{1}{Note.--- The complete version of this table is presented in the
electronic edition of the Journal.  A portion is shown here for guidance regarding
its content and format.}
\tablenotetext{a}{Units of 10$^{-16}$ erg/s/cm$^2$}
\end{deluxetable}

\end{document}